# Johnson-Lindenstrauss Compression with Neuroscience-Based Constraints*

Zeyuan Allen-Zhu
zeyuan@csail.mit.edu
MIT CSAIL

Rati Gelashvili
gelash@mit.edu
MIT CSAIL

Silvio Micali
silvio@csail.mit.edu
MIT CSAIL

Nir Shavit
shanir@csail.mit.edu
MIT CSAIL

## Abstract

Johnson-Lindenstrauss (JL) matrices implemented by sparse random synaptic connections are thought to be a prime candidate for how convergent pathways in the brain compress information. However, to date, there is no complete mathematical support for such implementations given the constraints of real neural tissue. The fact that neurons are either excitatory or inhibitory implies that every so implementable JL matrix must be *sign-consistent* (i.e., all entries in a single column must be either all non-negative or all non-positive), and the fact that any given neuron connects to a relatively small subset of other neurons implies that the JL matrix had better be *sparse*.

We construct sparse JL matrices that are sign-consistent, and prove that our construction is essentially optimal. Our work answers a mathematical question that was triggered by earlier work and is necessary to justify the existence of JL compression in the brain, and emphasizes that inhibition is crucial if neurons are to perform efficient, correlation-preserving compression.

## 1 Introduction

The existence of some form of compression in the brain is well accepted among neurobiologists. Its biological "evidence" proceeds from the brain's numerous convergent pathways, where information coming from a large number of neurons must be compressed into a small number of axons or neurons. Classical examples are the optic nerve fibers, that carry information about the activity of 100 times as many photoreceptors, [SD04] or the Pyramidal Tract fibers that carry information from the (orders-of-magnitude) larger motor cortex to the spinal cord [WDM12].

As far back as 1961, Barlow [Bar61] hypothesized that the role of early sensory neurons is to remove statistical redundancy in sensory input. This "efficient encoding" theory has been studied by many, as surveyed in depth by Simoncelli and Olshausen [SO01].

A recent survey by Ganguli and Sompolinsky [GS12] highlights the importance of compression and compressed sensing in the neural system for reducing the dimensionality of the activity pattern. A fundamental question they pose is "How much can a neural system reduce the dimensionality of its activity patterns without incurring a large loss in its ability to perform relevant computations?" They identify, as a minimal requirement, the importance of preserving the similarity structure of the neuronal representations at the source area, in order to capture the idea (see [KEMT07, RM04]) that in higher perceptual or association areas in the brain, semantically similar objects elicit similar neural activity patterns.

---

*A shorter version of this paper has appeared in the Proceedings of the National Academy of Sciences [AGMS14].

Ganguli and Sompolinsky suggest that such compression can be achieved in the brain via random synaptic-connectivity matrices implementing *Johnson-Lindenstrauss (JL)* matrices. However, since each neuron is either excitatory or inhibitory, an additional constraint, *sign-consistency*, is necessary for this implementation to work.

In this paper, we show for the first time that JL matrices can be simultaneously *compression-efficient*, *sparse* and *sign-consistent*, and are thus implementable by biologically plausible neural networks.

**JL Compression and Synaptic-Connectivity.** Informally, JL compression [JL84] uses a random matrix $A$ to map a long vector of reals, $x$, the *input*, to a much shorter vector of reals, $y = Ax$, the JL *output*. The JL result shows that if the number of input vectors one may ever need to compress is reasonably upperbounded, then the following property is satisfied:

*Inner-product Preservation:* $\quad \langle x, x' \rangle \approx \langle Ax, Ax' \rangle$ for all envisaged $x$ and $x'$.[1]

Notice that inner-product preservation implies the aforementioned "similarity property" of biological interest, that is

*Correlation Preservation:* $\quad \frac{\langle x,x' \rangle}{\|x\| \cdot \|x'\|} \approx \frac{\langle Ax, Ax' \rangle}{\|Ax\| \cdot \|Ax'\|}$ for all envisaged $x$ and $x'$.[2]

That is, similar JL inputs correspond to similar JL outputs.

The mentioned insight for implementing JL compression in the brain is random synaptic connectivity. An $m \times d$ JL matrix $A$ is biologically implemented via the synaptic connections among (the *axons* of) $d$ 'input' neurons and (the *dendrites* of) $m < d$ 'output' neurons. In essence,

> in a synaptic-connection matrix, the $j$th column corresponds to the connections of $j$th input neuron, $n_j$. An entry $(i,j)$ is 0 if $n_j$ does not connect to the $i$th output neuron; else, it is the strength of their synaptic connection.

The encouraging aspect of the above biological implementation of a JL matrix $A$ is that the random structure of $A$ matches the randomness of neural connections.[3]

## 1.1 Three Inter-Related Challenges

**Sign-Consistency.** Let us explain the constraint noted in [RAS10, reference 16]. According to Dale's principle, almost all neurons have one of the following two types: *excitatory* or *inhibitory*, but not both. The type of a neuron $n$ essentially determines the 'sign of the signal' it can transmit to a post-synaptic neuron $p$. As a standard approximation, an excitatory neuron $n$ can only increase the activity of $p$, and an inhibitory one can only decrease it. Thus, a synaptic-connection matrix must be *sign-consistent* ([RA06, GR09]). That is, the non-zero entries of a column $j$ must be either

(1) *all positive,* if the $j$th input neuron $n_j$ is excitatory, or

---

[1] As usual, $\langle x, x' \rangle$ represents the inner product of $x$ and $x'$, that is, $\sum_i x_i x'_i$. Inner-product preservation immediately implies (and is in fact equivalent to) *norm preservation*: namely, $\|x\| \approx \|Ax\|$ for all envisaged $x$.

[2] As usual, $\|x\|$ represents the $\ell_2$-norm of $x$ that is, $\sqrt{\langle x, x \rangle}$.

[3] While it may be easier to biologically construct a large random matrix, billions of years of evolution may not suffice for the emergence of a very special and very large matrix of neural connections. Moreover, this random construction need not be first found by Evolution, and then preserved genetically. That is, a good matrix $A$ need not be the same across different individuals of the same species. It suffices that our DNA ensures that each individual, during development, randomly constructs *his own* matrix $A$.

(2) *all negative*, if $n_j$ is inhibitory.

Unfortunately, typical JL matrices are *not* sign-consistent.

**Sparsity.** Let us emphasize another fundamental biological constraint: sparsity. A neuron may be connected to up to a few thousand postsynaptic neurons [KSJ00]. (Furthermore, two neurons typically share multiple connections.) Thus, no synaptic-connectivity matrix could implement a dense JL matrix when $m$ is large.[4]

As originally constructed, JL matrices were dense. Sparse JL matrices have been recent constructed ([Ach03, DKS10, KN10, BOR10, KN12]), but they are far from being sign-consistent. Therefore, although the sign consistency of synaptic action may have a few exceptions, the extent to which the above mathematical constructions may be biologically relevant is not clear.

**Efficiency.** As we have mentioned at the start of our introduction, implementing an $m \times d$ JL matrix in the brain is interesting only if $m$ is significantly smaller than $d$. (Of course, achieving such efficiency is more challenging with sign consistency, but Rajan and Abbott [RA06] have expressed optimism about the general ability to satisfy the latter constraint.)

**Three Prior Approaches.** Let us explain why these three challenges have not been simultaneously met.

A first and simplest way for JL matrices to be sign-consistent is for them not to have any negative entries, corresponding to synaptic-connectivity matrices without inhibitory neurons. However, it is not hard to prove that non-negative JL matrices must be *extremely* inefficient (e.g. $m \geq d/2$ for typical choices of parameters, see Appendix D.) This strong lower bound actually provides an additional evidence for the cruciality of inhibition for neural functions.

The result of Rajan and Abbott [RA06] on the eigenvalue spectra of square matrices implies a way to transform JL matrices into sign-consistent ones (subject to mild assumptions on the inputs). However, the sign-consistent JL matrices they obtained were very dense: half of their entries had to be non-zero. (See Section 2 for details.)

A third approach to sign-consistent JL matrices is implicitly provided by a transformation of Krahmer and Ward [KW11]. Indeed, when applied to non-negative restricted isometric property (RIP) matrices, their transformation yields sign-consistent JL matrices, but this construction can be proved to be much less efficient than ours (see Section 6).

## 1.2 Our Contributions

The mentioned biological constraints motivate the following purely mathematical question:

*How efficient can sparse (randomly constructed) and sign-consistent JL matrices be?*

We answer this question exactly by providing tight upper and lower bounds.

We begin by formally statement the classical JL lemma (using norms rather than inner products):

*Letting $m = \Theta(\varepsilon^{-2} \log(1/\delta))$, there exists a distribution $\mathcal{A}$ over $m \times d$ matrices such that, for any $x \in \mathbb{R}^d$, with probability at least $1 - \delta$:* $\quad \|Ax\|_2 = (1 \pm \varepsilon)\|x\|_2.$

The parameter $\varepsilon$ measures the *distortion* introduced by the JL compression; in particular, one may consider $\varepsilon = 10\%$ [GS12]. The parameter $\delta$ measures the *confidence* with which the $1 \pm \varepsilon$ distortion is guaranteed. Since one may only need to compress polynomially many (rather than exponentially many) vectors in his life-time, people typically choose $\delta = 1/\mathsf{poly}(d)$ and thus $\log(1/\delta) = O(\log d)$.

---

[4]Moreover, even if $m$ were small —e.g., $m = 1,000$— it seems hard to find in the brain a complete bipartite graph with $d$ 'inputs' and $m$ 'outputs' [Buz06, page 35].

(With this choice of $\delta$, after applying union bound, a matrix $A$ generated from $\mathcal{A}$ is capable of compressing $\mathsf{poly}(d)$ envisaged vectors from $\mathbb{R}^d$, with high confidence.)

We prove two main results:

**Thm 1.** "A CONSTRUCTION OF SPARSE, EFFICIENT, AND SIGN-CONSISTENT JL MATRICES." That is,

*Letting $m = \Theta(\varepsilon^{-2} \log^2(1/\delta))$, there exists a distribution $\mathcal{A}$ over $m \times d$ sign-consistent matrices such that, for any $x \in \mathbb{R}^d$, with probability at least $1 - \delta$: $\quad \|Ax\|_2 = (1 \pm \varepsilon)\|x\|_2$.*

More precisely, a matrix $A$ generated according to $\mathcal{A}$ enjoys the following properties.

- *Sparsity*: Each column has $\Theta(\varepsilon^{-1} \log(1/\delta))$ non-zero entries.
- *Same-Magnitude*: All non-zero entries have the same absolute value.
- *Simplicity*: The positions of the non-zero entries in a column, and the sign of a column itself, are both randomly selected, independent from other columns.

Note that Theorem 1 shows the norm preservation up to a multiplicative error $1 \pm \varepsilon$. This implies correlation preservation up to an additive error $\pm O(\varepsilon)$, that is,

$$\frac{\langle x, x' \rangle}{\|x\| \cdot \|x'\|} = \frac{\langle Ax, Ax' \rangle}{\|Ax\| \cdot \|Ax'\|} \pm O(\varepsilon) \ .$$

Note also that one cannot hope for a multiplicative error on correlation preservation because the correlation value is between $-1$ and $1$, and thus a multiplicative error would imply the ability of recovering orthogonal vectors (i.e., vectors with correlation zero) *precisely*.

**Thm 2.** "OUTPUT-LENGTH OPTIMALITY AMONG ALL SIGN-CONSISTENT JL MATRICES." That is,

*Let $\mathcal{A}$ be a distribution over $m \times d$ sign-consistent matrices such that, for any $x \in \mathbb{R}^d$, with probability at least $1-\delta$, $\|Ax\|_2 = (1\pm\varepsilon)\|x\|_2$. Then, $m = \tilde{\Omega}\left(\varepsilon^{-2} \log(1/\delta) \cdot \min\left\{\log d, \log(1/\delta)\right\}\right)$.*[5]

Note that in the interesting parameter regime of $\delta = 1/\mathsf{poly}(d)$, our lower bound becomes $\tilde{\Omega}(\varepsilon^{-2} \log^2(1/\delta))$, that is, it essentially matches our upper bound.

**Additional Results.** On the way to prove our main contribution, we derive two additional results that may be of independent interest. We defer the full statements and proofs to the relevant appendix sections.

**Thm 3.** Nelson and Nguyễn [NN13] prove that, in every JL matrix, there exists a column with large $\ell_0$-sparsity (i.e., with many non-zero entries). As part of our lower-bound proof, we *need* to strengthen their result by proving that, in every JL matrix, at least *half* of the columns have large $\ell_1$-sparsity (i.e., the sum of the entries' absolute values is large).

**Thm 4.** The techniques of our lower bound can also be used to prove tight bounds for non-negative (or sign-consistent) *restricted isometry property (RIP)* matrices —see Section 6. This simultaneously improves the previously known three lower bounds on $m$ —namely, $m = \Omega(\frac{k^2}{\varepsilon})$ [Cha08], $m = \Omega(\frac{k \log(d/k)}{\varepsilon})$ and $m = \Omega(\frac{k}{\varepsilon^2})$ [NN13]— to $m = \tilde{\Omega}\left(\frac{k^2 \log(d/k)}{\varepsilon^2}\right)$.

## 1.3 In Sum

Our work closes an open mathematical question that is necessary to justify the existence of JL compression in the brain. Our work provides the missing support by constructing JL matrices that are simultaneously sparse, sign-consistent and offer the most efficient JL compression possible; moreover, our work interestingly implies that inhibition is crucial if neurons are to perform efficient, correlation-preserving compression.

---

[5]Recall that the notation of $\tilde{\Omega}(N)$ signifies that logarithmic factors of $N$ are ignored. Thus, in our case, factors of $\log(1/\varepsilon)$ and $\log \log(1/\delta)$ are ignored in this lower bound.

Looking forward, the brain has inspired several models of computation, from perceptrons [Ros58] to neural networks [Fuk80], which have already proven fruitful in many a field, and in machine learning in particular.

Computer scientists have started studying computational models, that are increasingly biologically relevant, for fundamental tasks such as concept representation and formation [Val94], and memory allocation [Val05, FV09, Val12]. We consider our paper as a further step in this direction.

## 2 Related Mathematical Work

To the best of our knowledge, the only mathematical analysis of a random sign-consistency matrix is the one suggested by Rajan and Abbott [RA06], and followed by [GR09]. Although their results are about the eigenvalue spectra of a random sign-consistent *square* matrix, it implies[6] the following way of constructing an $m \times d$ sign-consistent JL matrix $A'$.

- First, construct an $m \times d$ JL matrix $A$, by randomly assigning each entry of $A$ from $\{-1/\sqrt{m}, 1/\sqrt{m}\}$.
- Second, construct an $m \times d$ special matrix $M$, by assigning each entry of (a random) half of the columns of $M$ to be $-1/\sqrt{m}$, and each entry of the remaining half to be $1/\sqrt{m}$.
- At last, set $A' \stackrel{\text{def}}{=} \frac{1}{\sqrt{2}}(A + M)$.

Then, $A'$ is sign-consistent, and $A'x = Ax$ (so $A'$ is JL) assuming that $x$ satisfies $\sum_{i \in [d]} x_i = 0$. This assumption aside, however, the resulting matrix $A'$ must be very dense.

The classical JL construction requires a distribution over *dense* matrices (e.g., i.i.d. Gaussian or Rademacher entries), but achieves a target dimension of $m = O(\varepsilon^{-2} \log(1/\delta))$ which is essentially optimal [Alo09]. A beautiful line of work [AL08, AC09, KW11, HV11, Vyb11, AL13] has made use of the Hadamard or Fourier matrices in the JL construction, to speed up the matrix-vector multiplication to nearly-linear time. However, their matrices are *dense too*. Recent constructions ([Ach03, DKS10, KN10, BOR10, KN12]) yield sparse JL matrices that have $O(\varepsilon^{-1} \log(1/\delta))$ rows, which have been shown to be essentially optimal [NN13].

Although not applicable to JL matrices, Clarkson and Woodruff have shown how to construct sign-consistent and *optimally sparse* (namely, a single non-zero entry per column) random matrices [CW13]. Their matrices preserve correlation for inputs satisfying an algebraic constraint, namely, coming from a *hidden subspace*. By contrast, we want to compress arbitrary inputs.

For numerous of applications of JL compression in computer science, see [Ind01, Vem04].

A JL matrix $A$ can be easily constructed (with very high probability) by choosing each entry at random. Of course, given such a randomly constructed matrix, it would be nice to reconstruct, with meaningful approximation, the original JL-compressed input $x$ from $Ax$;[7] but this cannot be done without assuming that the inputs are of a *restricted type* (e.g., close to vectors with few non-zero entries [CT05, CRT06]). Yet, even without reconstructing the inputs, inner-product preservation allows one to perform a variety of fundamental computations on the JL outputs directly, and thus with great efficiency. This includes nearest neighbors [IM98], classification [Blu05], regression [ZLW09], and many others.

---

[6]In fact, given any random matrix whose eigenvalues are randomly distributed on the complex unit disk, a random subset of its rows forms a JL matrix.

[7]To be sure, inner-product preservation always implies a weak form of reconstructability. Namely, each entry $x_i$ of an input vector $x$ can be reconstructed up to an additive error of $\varepsilon \cdot \|x\|_2$.

## 3 A Simple Experimental Illustration

Let us consider a simple experiment to numerically verify the dependency $m = \Theta(\varepsilon^{-2}\log(1/\delta))$ in the classical JL construction and the dependency $m = \Theta(\varepsilon^{-2}\log^2(1/\delta))$ in our new construction. Rather than fixing the distortion $\varepsilon$ and the confidence $\delta$, and computing the target dimension $m$, we find it more convenient to fix $m$ and $\delta$ first, and then compute $\varepsilon$.

Specifically, we fix $d = 3000$ and $\delta = 0.1$; consider the following values of $m$: $m = 10, 20, 40, \ldots, 1280$; and then numerically compute the distortion $\varepsilon$ for each value of $m$, both for the classical and our new JL construction.

In the classical JL construction, the $m \times d$ dimension matrix $A$ is chosen so that each entry is either $1/\sqrt{m}$ or $-1/\sqrt{m}$, each with half probability.

In our new construction, we first define the column sparsity $s = \lfloor\sqrt{m}\rfloor$. (Note that this is consistent with the parameters suggested by Theorem 1: $s = \Theta(\varepsilon^{-1}\log(1/\delta)) = \Theta(\sqrt{m})$.) Then, for each column of $A$, we randomly choose $s$ entries of this column, and then flip a fair coin: if Heads, we set each of these $s$ entries to $1/\sqrt{s}$; if Tails, we set each of them to $-1/\sqrt{s}$.

For the above two constructions, we apply the JL transformation $Ax$ on 1000 randomly chosen inputs $x_1, \ldots, x_{1000} \in \mathbb{R}^d$. For each construction, we compute the 1000 distortions $\left|\frac{\|Ax_i\|_2}{\|x_i\|_2} - 1\right|$, call $\varepsilon$ the 100-th highest distortion, and plot $\varepsilon$ in Figure 1. (This process is equivalent to choosing $\delta = 0.1$ and throwing out the highest $\delta$-fraction of the distortions. Indeed, $100 = \delta \cdot 1000$.)

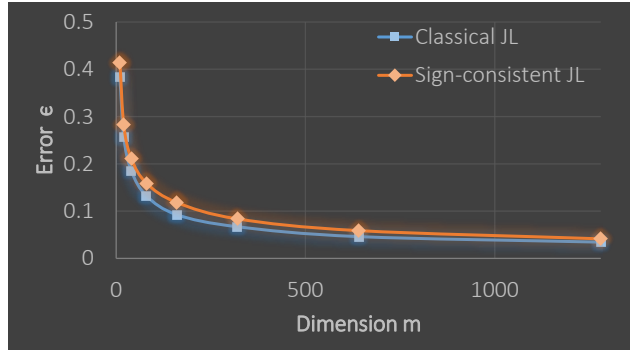

Figure 1: The classical and our sign-consistent JL constructions.

The experiment illustrates that for both curves, whenever $m$ is enlarged by a factor of 4, the error $\varepsilon$ decreases approximately by a factor of 2. This corresponds to the dependency $m \propto \varepsilon^{-2}$ in both constructions. Also note that the blue curve falls slightly below the red curve, corresponding to the difference between the dimension choice of $m = \Theta(\varepsilon^{-2}\log(1/\delta))$ in the classical JL construction, and $m = \Theta(\varepsilon^{-2}\log(1/\delta)^2)$ in ours.

## 4 Proof Sketch of Theorem 1

Let $\mathcal{A}^{m,d,s}$ be the distribution of $m \times d$ matrices defined as follows. For each of the $d$ columns, we choose uniformly at random $s$ distinct entries (out of $\binom{m}{s}$ possibilities), and assign a random value between $\{-1/\sqrt{s}, 1/\sqrt{s}\}$ (with half probability each) to these $s$ entries, while leaving it zero in other entries of the same column.[8]

---

[8]Our theorem remains true if one divides each column into $\lceil m/s \rceil$ blocks and chooses one random entry from each block; and/or if one uses $\Theta(\log(1/\delta))$-wise independent hash functions to generate $\mathcal{A}^{m,d,s}$.

**Theorem 1.** *Letting $m = \Theta(\varepsilon^{-2} \log^2(1/\delta))$ and $s = \Theta(\varepsilon^{-1} \log(1/\delta))$, for any $x \in \mathbb{R}^d$, with probability at least $1 - \delta$, we have $\|Ax\|_2 = (1 \pm \varepsilon)\|x\|_2$ over the choice of $A \sim \mathcal{A}^{m,d,s}$.*

The proof of Theorem 1 is quite complex and is given in Appendix A.
(In particular, the classical technique of the Hanson-Wright inequality fails to give a tight upper bound in our case, just like [KN12].) Below, we just outline the important ingredients of the proof.

**Proof sketch.** Observe that, the entries of a matrix $A \in \mathbb{R}^{m \times d}$ that we construct can be written as $A_{i,j} = \eta_{i,j}\sigma_j/\sqrt{s}$, where $\sigma_j \in \{-1, 1\}$ is chosen uniformly at random, and $\eta_{i,j} \in \{0, 1\}$ is an indicator variable for the event $A_{i,j} \neq 0$. All the $\{\sigma_j\}_{j \in [d]}$ are independent; $\{\eta_{i,j}\}_{i \in [m], j \in [d]}$ are independent across columns, but not independent (and in fact negatively correlated) in the same column, since there are exactly $s$ non-zero entries per column.

Given any fixed $x \in \mathbb{R}^d$ with $\|x\|_2 = 1$, let us study the following random variable :

$$Z \stackrel{\text{def}}{=} \|Ax\|_2^2 - 1 = \frac{1}{s} \cdot \sum_{r=1}^{m} \sum_{i \neq j \in [d]} \eta_{r,i}\eta_{r,j}\sigma_i\sigma_j x_i x_j,$$

To show that $|Z| \leq \varepsilon$ with probability at least $1 - \delta$, we need a good upper bound on the $t$-th moment of $Z$ (notice that we will eventually choose $t = \Theta(\log(1/\delta))$):

$$s^t \cdot \mathbb{E}[Z^t] = \sum_{\substack{i_1,\ldots,i_t,j_1,\ldots,j_t \in [d] \\ i_1 \neq j_1, \ldots, i_t \neq j_t}} \left(\prod_{u=1}^{t} x_{i_u} x_{j_u}\right) \left(\mathbb{E}_\sigma \prod_{u=1}^{t} \sigma_{i_u} \sigma_{j_u}\right) \left(\mathbb{E}_\eta \prod_{u=1}^{t} \sum_{r=1}^{m} \eta_{r,i_u} \eta_{r,j_u}\right)$$

In [KN12], they analyzed a similar expression but with $\sigma_{i_u}\sigma_{j_u}$ replaced by $\sigma_{r,i_u}\sigma_{r,j_u}$. In their case, they decompose $Z$ into sub-expressions $Z = Z_1 + \cdots + Z_m$: each $Z_r$ contains all terms with the same row $r$ (e.g. $\sigma_{r,\star}$ and $\eta_{r,\star}$), and can be analyzed separately. This greatly simplifies their job because $Z_r$ and $Z_{r'}$ are negatively correlated when $r \neq r'$. In contrast, if we did the same thing, we would not have the same negative correlation anymore: $Z_r$ and $Z_{r'}$ both contain the same random variables $\sigma_i$ for all $i \in [d]$. Thus, we have to analyze the whole expression at once.

Reusing ideas from [KN10, KN12, BOR10], we analyze $Z^t$ by associating monomials that appear in $Z^t$ to *directed multigraphs with labeled edges*: an $x_{i_u} x_{j_u}$ term corresponds to a directed edge with label $u$ from vertex $i_u$ to vertex $j_u$. We then group the monomials together based on their associated graphs, and prove the following lemma. (Its proof is analogous to [KN12, (13)] but is more tedious.)

**Lemma 4.1.** $s^t \cdot \mathbb{E}[Z^t] \leq e^t \sum_{v=2}^{t} \sum_{G \in \mathcal{G}''_{v,t}} \left(\frac{1}{t^t} \prod_{p=1}^{v} \sqrt{d_p}^{d_p}\right) \cdot \sum_{r_1,\ldots,r_t \in [m]} \prod_{i=1}^{w} \left(\frac{s}{m}\right)^{v_i}$. *Here,*

- $\mathcal{G}''_{v,t}$ *is a set of directed multigraphs with $v$ labeled vertices (1 to $v$) and $t$ labeled edges (1 to $t$).*
- $d_p$ *is the total degree of vertex $p \in [v]$ in a graph $G \in \mathcal{G}''_{v,t}$.*[9]
- *$w$ and $v_1, \ldots, v_w$ are defined by $G$ and $r_1, \ldots, r_t$ as follows. Let an edge $u \in [t]$ be colored with $r_u \in [m]$, then we define $w$ to be the number of distinct colors used in $r_1, \ldots, r_t$, and $v_i$ to be the number of vertices incident to an edge with color $i \in [w]$.*

As one may have observed, for the aforementioned reason, we need to deal with many rows (e.g., row $r_1, \ldots, r_t$) together, introducing a concept of *color* defined above. To be precise, a directed edge $(i_u, j_u)$ is now also colored with $r_u \in [m]$, and this is a major difference between our Lemma 4.1 and for instance [KN12, (13)]. In essence, we are dealing with 3-dimensional tuples $(i_u, j_u, r_u)$ rather than just $(i_u, j_u)$.

---

[9]The total degree of a vertex is defined as the number of incident edges regardless of direction.

This difference is critical for obtaining a tight bound for $Z^t$: one has to bound the $\prod_{p=1}^{v} \sqrt{d_p}^{d_p}$ terms separately for graphs of different colors (as otherwise he will lose a $\log(1/\delta)$ factor in the proof). In other words, instead of enumerating $G \in \mathcal{G}''_{v,t}$ as a whole, we now have to enumerate subgraphs of different colors separately, and then combine the results. Below is one way (and perhaps the only way the authors believe) to enumerate $G$ that can lead to tight upper bounds

$$s^t \cdot \mathbb{E}[Z^t] \leq e^t \underbrace{\sum_{v=2}^{t}}_{\text{i}} \underbrace{\sum_{w=1}^{t} \binom{m}{w}}_{\text{ii}} \underbrace{\sum_{\substack{c_1,\ldots,c_w \\ c_1+\ldots+c_w=t \\ c_i \geq 1}} \binom{t}{c_1,\ldots,c_w}}_{\text{iii}} \underbrace{\sum_{\substack{v_1,\ldots,v_w \\ 2 \leq v_i \leq 2c_i}} \left(\frac{s}{m}\right)^{v_1+\cdots+v_w}}_{\text{iv}} \underbrace{\sum_{f_1,\ldots,f_w}}_{\text{v}} \underbrace{\sum_{\forall i, G_i \in \mathcal{G}''_{v_i,c_i}}}_{\text{vi}} \frac{1}{t^t} \prod_{p=1}^{v} \sqrt{d_p}^{d_p} \quad (4.1)$$

This gigantic expression enumerates all $G \in \mathcal{G}''_{v,t}$ and their colorings $r_1, \ldots, r_t \in [m]$ in six steps:

(i). Number of graph vertices, $v \in \{2, \ldots, t\}$; the vertices are labelled by $1, 2, \ldots, v$.

(ii). Number of used edge colors, $w \in \{1, \ldots, t\}$, and all $\binom{m}{w}$ possibilities of choosing $w$ colors.

(iii). Edge colorings of the graph using selected $w$ colors: how many (denoted by $c_i \geq 1$) edges are colored in color $i$ and which of the $t$ edges are colored in color $i$.

(iv). Number of vertices $v_i \in \{2, \ldots, 2c_i\}$ in each $G_i$, the subgraph containing edges of color $i$.

(v). All possible increasing functions $f_i : [v_i] \to [v]$, such that $f_i(j)$ maps vertex $j$ in $G_i$ to the $f_i(j)$-th global vertex. (And we ensure $f_i(j) < f_i(k)$ for $j < k$ to reduce double counting.)

(vi). All graphs $G_i \in \mathcal{G}''_{v_i,c_i}$ with $v_i$ labeled vertices (1 to $v_i$) and $c_i$ labeled edges (1 to $c_i$).

(Using all the information above, $d_p$, the degree of vertex $p \in [v]$ is well defined.)

We emphasize here that any pair of graph $G \in \mathcal{G}''_{v,t}$ and coloring $r_1, \ldots, r_t \in [m]$ will be generated *at least once* in the above procedure.[10] Thus, (4.1) follows from Lemma 4.1, since the summation terms also have the same value $\left(\frac{s}{m}\right)^{v_1+\cdots+v_w} \frac{1}{t^t} \prod_{p=1}^{v} \sqrt{d_p}^{d_p}$.

It is now possible to consider $G_i$'s separately in (4.1) and prove the following lemma:

**Lemma 4.2.**

$$s^t \cdot \mathbb{E}[Z^t] \leq 2^{O(t)} \sum_{v=2}^{t} \sum_{w=1}^{t} \binom{m}{w} \sum_{\substack{c_1,\ldots,c_w \\ c_1+\ldots+c_w=t \\ c_i \geq 1}} \binom{t}{c_1,\ldots,c_w} \sum_{\substack{v_1,\ldots,v_w \\ 2 \leq v_i \leq 2c_i}} \prod_{j=1}^{w} \left(\frac{s}{m}\right)^{v_j} v_j^{c_j} \binom{v-1}{v_j-1} .$$

Some delicate issues arise here. For instance, one may use the Cauchy-Shwartz technique of [KN12] to deduce

$$\sum_{f_1,\ldots,f_w} \sum_{\forall i, G_i \in \mathcal{G}''_{v_i,c_i}} \frac{1}{t^t} \prod_{p=1}^{v} \sqrt{d_p}^{d_p} \leq \sum_{\substack{v_1,\ldots,v_w \\ 2 \leq v_i \leq 2c_i}} \prod_{j=1}^{w} v_j^{c_j} \binom{v}{v_j} ,$$

getting a weaker upper bound as it replaces $\binom{v-1}{v_j-1}$ with $\binom{v}{v_j}$ in Lemma 4.2. However, even such a simple replacement leads to a $\log(1/\delta)$ factor loss! At last, after enduring layers of algebraic simplifications we prove

**Lemma 4.3.** $s^t \cdot \mathbb{E}[Z^t] \leq 2^{O(t)} \cdot t^t \left(\frac{s^2}{m}\right)^t$

---

[10]This follows from the fact that $G$ and $r_1, \ldots, r_t$ together determine (a) $w$, the number of used colors, (b) $G_i$ for each $i \in [w]$ (with $v_i$ vertices and $c_i$ edges), the subgraph of $G$ of the $i$-th used color, and (c) $f_i$, the vertex mapping from $G_i$ back to $G$. Any such triple will be generated at least once in (4.1). Note also, we may have double counts but it will not affect our asymptotic upper bound.

By Lemma 4.3, there exist constant $C$ such that $\mathbb{E}[Z^t] \leq \left(\frac{Cts}{m}\right)^t$. Using Markov's inequality, we have $\Pr[|Z| > \varepsilon] \leq \frac{\mathbb{E}[Z^t]}{\varepsilon^t} \leq \left(\frac{Cts}{\varepsilon m}\right)^t$. We now set parameters: $t \stackrel{\text{def}}{=} \log \frac{1}{\delta}$, $s \stackrel{\text{def}}{=} \varepsilon^{-1} t$ and $m \stackrel{\text{def}}{=} \frac{\varepsilon^{-1} ts}{2C}$. Plugging them in we get $\Pr[|Z| > \varepsilon] \leq \delta$ as desired, finishing the proof of Theorem 1. ∎

## 5 Proof of Theorem 2

Here we formally state and prove our second theorem.

**Theorem 2.** *There is some fixed $\varepsilon_0 \in (0, 1/2)$ such that*
*For all $\varepsilon \in (1/\sqrt{d}, \varepsilon_0)$, all $m \leq O(d/\log(1/\varepsilon))$, and all $\delta \leq \varepsilon^{12}$, the following holds.*
*Let $\mathcal{A}$ be a distribution over $m \times d$ sign-consistent matrices such that, for any $x \in \mathbb{R}^d$, with probability at least $1 - \delta$, the $\ell_2$ embedding $\|Ax\|_2 = (1 \pm \varepsilon)\|x\|_2$ has $\varepsilon$-distortion. Then,*
$$m = \Omega\left(\frac{\varepsilon^{-2} \log(1/\delta)}{\log\left(\varepsilon^{-2} \log(1/\delta)\right)} \min\left\{\log d, \log(1/\delta)\right\}\right)$$

### 5.1 Strengthening the Sparsity Lower Bound of [NN13]

We begin with a simple fact connecting JL matrices to $\varepsilon$-*incoherence* matrices. Given a matrix $A \in \mathbb{R}^{m \times n}$, let us denote its columns by $v_1, \ldots, v_n \in \mathbb{R}^m$. $A$ is said to be $\varepsilon$-incoherent if for all $i \neq j$, $|\langle v_i, v_j \rangle| \leq \varepsilon$, and for all $i$, $\|v_i\|_2 = 1$. Then,

**Fact 5.1** ([Alo09]). *Let $\{e_1, \ldots, e_n\}$ be the first $n$ unit basis vectors of $\mathbb{R}^d$ where $n \in [d]$. Given any matrix $A \in \mathbb{R}^{m \times d}$ satisfying for any $x, y \in \{0, e_1, \ldots, e_n\}$, $\|A(x - y)\|_2 = (1 \pm \varepsilon)\|x - y\|_2$, we have that the first $n$ columns of $A$ (after normalization) form an $O(\varepsilon)$-incoherent submatrix.*

Owing to this fact above, lower bounds on $\varepsilon$-incoherent matrices directly translate to that of JL matrices, after choosing appropriate values of $n$ (and we will eventually choose $n = \min\{d, \frac{1}{\delta^{1/4}}\}$). In particular, Nelson and Nguyễn [NN13] show that in an $\varepsilon$-incoherent matrix, there exists at least some column whose $\ell_0$-sparsity —i.e., number of non-zero entries— is $\Omega(\varepsilon^{-1} \log n / \log(m/\log n))$.

We prove a strengthened version of this $\ell_0$-sparsity lower bound. Namely, we show a lower bound on the $\ell_1$ norm (which implies the same lower bound on the $\ell_0$-sparsity), on at least half of the columns of $A$ rather than a single column. More precisely, we show that (whose proof is deferred to Appendix B):

**Theorem 3.** *There is some fixed $0 < \varepsilon_0 < 1/2$ so that the following holds. For any $1/\sqrt{n} < \varepsilon < \varepsilon_0$ and $m < O(n/\log(1/\varepsilon))$, let $A \in \mathbb{R}^{m \times n}$ be an $\varepsilon$-incoherent matrix. Then, at least half of the columns $A$ must have $\ell_1$ norm being $\Omega(\sqrt{\varepsilon^{-1} \log n / \log(m/\log n)})$.*

It is worth noting that our strengthened lower bound implies: (1) the average $\ell_1$ norm of the columns of $A$ is $\Omega(\sqrt{\varepsilon^{-1} \log n / \log(m/\log n)})$, (2) at least half of the columns of $A$ must have $\ell_0$-sparsity $\Omega(\varepsilon^{-1} \log n / \log(m/\log n))$.

### 5.2 Dimension Lower Bound for Sign-Consistent JL Matrices

The lower bound in Section 5.1 works as follows. There is a fixed hard instance of vectors, i.e., $\{0, e_1, \ldots, e_n\}$, so that *even if* the adversary knows this hard instance, he cannot produce a good $\varepsilon$-incoherent matrix (and thus a JL matrix), unless the sparsity reaches the desired lower bound.

In this section, we lower bound $m$ in a conceptually different way. We will choose the hard instance *after* the JL construction $\mathcal{A}$ (i.e., the distribution of the matrices) is determined, and then show that $\mathcal{A}$ must perform bad on this hard instance, unless $m$ is large. This is a major difference between our proof and the related lower bounds for JL matrices, see instance [Alo09, NN13].

**High Level Proof Sketch.** Let us assume for simplicity that $\delta = 1/\mathsf{poly}(d)$ and $n = d$; the general case needs to be done more carefully. Take an arbitrary distribution $\mathcal{A}$ of $m \times n$ matrices satisfying the JL property with $\varepsilon$ and $\delta$. We divide our proof into three steps.

- In the first step, we use our Theorem 3 to conclude that almost all $A$ drawn from $\mathcal{A}$ (being $\varepsilon$-incoherent) must have an average $\ell_1$-sparsity (over the columns) $\geq \sqrt{s}$, where $s = \tilde{\Omega}(\varepsilon^{-1} \log n)$. For simplicity, assume that all matrices $A \sim \mathcal{A}$ have such property.

- In the second step, we use this $\ell_1$-sparsity lower bound on $A \sim \mathcal{A}$ to deduce that $A$ must have a large pairwise *column correlation*. Namely, $\sum_{i,j} |\langle v_i, v_j \rangle| \geq sn^2/m$ where $v_i$ represents the $i$-th column of $A$. By an averaging argument, we can pick some subset $S \subset [d]$ of the columns where $|S| = N \stackrel{\text{def}}{=} 1/\log n$, such that the correlations between columns in $S$ are also large: namely, $\sum_{i,j \in S} |\langle v_i, v_j \rangle| \geq \Omega(sN^2/m)$. This is formally proved as Lemma 5.2 below.

- In the third step, we begin with a wishful thinking. By the property of JL, $A$ must satisfy $\|\sum_i v_i\|_2^2 = (1 \pm \varepsilon)^2 N$ because $A$ must preserve the $\ell_2$-norm on vector $x = \sum_{i \in S} e_i$. If all columns $v_i$ for $i \in S$ *had* positive signs, then $\|\sum_i v_i\|_2^2 = N + \sum_{i,j \in S} \langle v_i, v_j \rangle \leq N + \varepsilon N$. This formula, when combined with the previous step of $\sum_{i,j \in S} |\langle v_i, v_j \rangle| \geq \Omega(sN^2/m)$, would give $sN^2/m \leq \varepsilon N$ so we have $m \geq \varepsilon^{-1} \log n \cdot s = \tilde{\Omega}(\varepsilon^{-2} \log^2 n)$ and we *would* be done.

  To fix this, we need to construct the hard instance more carefully. Instead of letting a single vector $x$ be the hard instance, we want all the $2^N$ possible combinations $X = \{\sum_{i \in S} s_i e_i : s_i \in \{-1, 1\}\}$ to present in the hard instance. We can afford this since $2^N = n$. Therefore, although the the signs of the columns in $S$ in a matrix $A$ may vary as $A \sim \mathcal{A}$, there is always some $x \in X$ that makes all the correlation to go positive, and the above sign issue goes away.

In sum, our hard instance so constructed depends on $S$, a set chosen *after* we see the distribution $\mathcal{A}$; and it contains $\mathsf{poly}(n)$ vectors. We now begin with our averaging lemma for the second step.

**Lemma 5.2.** *For any distribution of $m \times n$ matrices $\mathcal{A}$ such that (1) $m < n/2$, (2) each column of $A \in \mathcal{A}$ is normalized and (3) $\mathbb{E}_{A \sim \mathcal{A}} \left[\frac{1}{n} \sum_{i,j} |A_{i,j}|\right] = \sqrt{s}$, there exist a subset $S \subseteq [n]$ of columns with cardinality $|S| = N$ (for any $N \in [n]$) such that*

$$\mathbb{E}_{A \sim \mathcal{A}}\left[\sum_{i,j \in S, i \neq j} |\langle v_i, v_j \rangle|\right] \geq \Omega(sN^2/m) \ .$$

*Here, as usual, we denote by $v_i$ the $i$-th column of $A$.*

*Proof.* We compute this quantity via an averaging argument. On one hand, for a matrix $A$:

$$\sum_{i,j \in [n], i \neq j} |\langle v_i, v_j \rangle| = \sum_{r=1}^m \left(\left(\sum_{i \in [n]} |A_{r,i}|\right)^2 - \sum_{i \in [n]} A_{r,i}^2\right) = \sum_{r=1}^m \left(\sum_{i \in [n]} |A_{r,i}|\right)^2 - n \geq \frac{1}{m}\left(\sum_{r,i} |A_{r,i}|\right)^2 - n$$

and therefore when taking over the distribution of $A \sim \mathcal{A}$ we have

$$\mathbb{E}_{A \sim \mathcal{A}}\left[\sum_{i,j \in [n], i \neq j} |\langle v_i, v_j \rangle|\right] \geq \mathbb{E}_{A \sim \mathcal{A}}\left(\frac{1}{m}\left(\sum_{r,i} |A_{r,i}|\right)^2 - n\right) \geq \frac{1}{m}\left(\mathbb{E}_{A \sim \mathcal{A}} \sum_{r,i} |A_{r,i}|\right)^2 - n = sn^2/m - n = \Omega(sn^2/m) \ .$$

On the other hand, we note that

$$\sum_{S \subset [n], |S| = N} \sum_{i,j \in S, i \neq j} |\langle v_i, v_j \rangle| = \binom{n-2}{N-2} \cdot \sum_{i,j \in [n], i \neq j} |\langle v_i, v_j \rangle|$$

and there are a total number of $\binom{n}{N}$ subsets $S$ of cardinality $N$. By an averaging argument, there exist some subset $S^* \subset [n]$ satisfying

$$\mathop{\mathbb{E}}_{A \sim \mathcal{A}}\left[\sum_{i,j \in S^*, i \neq j} |\langle v_i, v_j \rangle|\right] \geq \frac{1}{\binom{n}{N}}\binom{n-2}{N-2} \cdot \mathop{\mathbb{E}}_{A \sim \mathcal{A}}\left[\sum_{i,j \in [n], i \neq j} |\langle v_i, v_j \rangle|\right] \geq \Omega(sN^2/m) \; . \quad \square$$

**Proof of Theorem 2.** We are now ready to implement the aforementioned high level proof sketch. Given any such distribution $\mathcal{A}$, we let $n = \min\{d, \frac{1}{\delta^{1/4}}\}$. Using union bound, with probability at least $1 - O(\delta n^2) \geq 1 - O(\frac{1}{n^2})$, a matrix $A$ drawn from $\mathcal{A}$ will preserve $\ell_2$ norms with $\varepsilon$ distortion for all vector $x = v_1 - v_2$ where $v_1, v_2 \in \{0, e_1, \ldots, e_n\}$.

In other words, owing to Fact 5.1, with probability at least $1 - O(\delta n^2) \geq 1 - O(\frac{1}{n^2})$, a matrix $A$ drawn from $\mathcal{A}$ satisfies that its first $n$ columns form an $O(\varepsilon)$-incoherent $m \times n$ submatrix (after column normalizations). Let $\mathcal{A}'$ be this subdistribution of $m \times n$ $O(\varepsilon)$-incoherent matrices.

Thanks to our strengthened Theorem 3 on the $\ell_1$-sparsity,[11] letting $\sqrt{s} \stackrel{\text{def}}{=} \mathbb{E}_{A' \sim \mathcal{A}'}\left[\frac{1}{n}\sum_{i,j} |A'_{i,j}|\right]$, we must have $s = \Omega(\varepsilon^{-1} \log n / \log(m/\log n))$. We plug this distribution $\mathcal{A}'$ into Lemma 5.2 along with the choice of $N \stackrel{\text{def}}{=} \log(1/\delta^{1/2})$, and deduce that

$$\mathop{\mathbb{E}}_{A' \sim \mathcal{A}'}\left[\sum_{i,j \in S, i \neq j} |\langle v_i, v_j \rangle|\right] \geq \Omega(sN^2/m) \; . \tag{5.1}$$

Now comes the important construction. Let us study define the following set of $2^N$ vectors,

$$X = \left\{\sum_{i \in S} s_i e_i \; : \; \forall i \; s_i \in \{-1, 1\}\right\} \subset \mathbb{R}^n \subset \mathbb{R}^d \; .$$

Because $1 - O(\delta 2^N) \geq 1 - O(\frac{1}{n^2})$, with probability at least $1 - O(\frac{1}{n^2})$, all vectors in $x \in X$ must have their $\ell_2$ norm preserved within $\varepsilon$-distortion over the choice of $A \sim \mathcal{A}$. This is also true with probability at least $1 - O(\frac{1}{n^2})$ over the choice of $A' \sim \mathcal{A}'$ by union bound. Let us denote by $\mathcal{A}''$ this subdistribution of matrices $A' \sim \mathcal{A}'$ such that

$$\forall x \in X, \quad \|A'x\|_2 = (1 \pm O(\varepsilon))\|x\|_2 \; .$$

By the above argument, $\mathcal{A}''$ contributes to at least $1 - O(\frac{1}{n^2})$ probability mass in both $\mathcal{A}'$ and $\mathcal{A}$.

Next, for each matrix $A'' \in \mathcal{A}''$, we claim that $\sum_{i,j \in S, i \neq j} |\langle v_i, v_j \rangle| = O(\varepsilon N)$. This is because, letting $x = \sum_{i \in S} s_i e_i \in X$ be a vector where $s_i$ coincides with the column sign of $v_i$, then

$$O(\varepsilon N) \geq \|A''x\|_2^2 - \|x\|_2^2 = \|\sum_{i \in S} s_i v_i\|_2^2 - \|\sum_{i \in S} e_i\|_2^2 = \sum_{i,j \in S, i \neq j} |\langle v_i, v_j \rangle| \; .$$

Therefore, we must have

$$\mathop{\mathbb{E}}_{A' \sim \mathcal{A}'}\left[\sum_{i,j \in S, i \neq j} |\langle v_i, v_j \rangle|\right] \leq \mathop{\mathbb{E}}_{A'' \sim \mathcal{A}''}\left[\sum_{i,j \in S, i \neq j} |\langle v_i, v_j \rangle|\right] + O\left(\frac{1}{n^2}\right)N^2 \leq O\left(\varepsilon N + \frac{N^2}{n^2}\right)$$

---

[11]To be precise, we need to verify that $1/\sqrt{n} < \varepsilon$. This is easy given our assumption of $1/\sqrt{d} < \varepsilon$ and $\delta < \varepsilon^{12}$. In addition, we need to verify that $m < O(n/\log(1/\varepsilon)) = O\big(\min\{d, 1/\delta^{1/4}\}/\log(1/\varepsilon)\big)$. The first term in min is true by assumption: $m \leq O(d/\log(1/\varepsilon))$. For the second term, suppose it is false, then we get $m \geq \Omega(\frac{1}{\delta^{1/4}}/\log(1/\varepsilon)) \geq \Omega(\frac{\varepsilon^{-2.5}}{\log(1/\varepsilon)} \cdot \frac{1}{\delta^{1/24}}) \geq \Omega(\varepsilon^{-2} \log^2(1/\delta))$, using $\delta \leq \varepsilon^{12}$ and $\delta$ and $\varepsilon$ smaller than some sufficiently small constant.

when comparing the above lower bound and (5.1), we get $m = \Omega(\varepsilon^{-1} N \cdot s) = \Omega(\varepsilon^{-1} \log(1/\delta) s)$. Substituting the $\ell_1$-sparsity lower bound for $s$ we have

$$m \geq \Omega(\varepsilon^{-2} \log(1/\delta) \log n / \log (m / \log n)) \implies$$
$$m \geq \Omega(\varepsilon^{-2} \log(1/\delta) \log n / \log (\varepsilon^{-2} \log(1/\delta)))$$
$$= \Omega\Big(\frac{\varepsilon^{-2} \log(1/\delta)}{\log (\varepsilon^{-2} \log(1/\delta))} \min\big\{\log d, \log(1/\delta)\big\}\Big) \ . \qquad \blacksquare$$

## 6 Tight bounds for Non-Negative RIP Matrices

Let us recall the definition of the closely related *restricted isometry property (RIP)* matrices.

**Definition 6.1** ([CT05, CRT06]). *Let $k \in [d]$ be a positive integer, a matrix $A \in \mathbb{R}^{m \times d}$ is said to satisfy the $k$-restricted isometry property with distortion $\varepsilon$, or $(k, \varepsilon)$-RIP for short, if for all $x \in \mathbb{R}^d$ with at most $k$ non-zero entries: $(1 - \varepsilon) \|x\|_2 \leq \|Ax\|_2 \leq (1 + \varepsilon) \|x\|_2$ .*

*(WLOG, in this paper we assume that each column of $A$ has $\ell_2$ norm precisely equal to 1.)*

RIP matrices have a tremendous number of applications in compressed sensing [Don06]. Indeed, the works [CT05, CRT06] show that RIP matrices allow the approximate reconstruction (via linear programming) of an original sparse input $x$ given $Ax$.

Typically, from any JL construction with $\varepsilon$ distortion and $\delta$ confidence, one can pick $\log(1/\delta) = O(k \log(d/k))$ and perform a suitable union bound to get an $(k, O(\varepsilon))$-RIP matrix with dimension $m = O(\varepsilon^{-2} k \log(d/k))$. See for instance [BDDW08] for a simple proof.

In fact, Krahmer and Ward have shown a weak converse:

**Theorem 6.2** ([KW11]). *Given an $(k, \varepsilon)$-RIP matrix $A \in \mathbb{R}^{m \times d}$ satisfying $k = \Omega(\log(1/\delta))$, and let $A'$ be $A$ but after randomizing its column signs[12] Then, for any $x \in \mathbb{R}^d$, we have $(1 - \varepsilon) \|x\|_2 \leq \|A'x\|_2 \leq (1 + \varepsilon) \|x\|_2$ with probability at least $1 - \delta$.*

**Constructing Sign-Consistent JL Matrices From Krahmer-Ward.** To the best of our knowledge, the only known way to construct a sparse and sign-consistent JL matrix, is to start from a sparse and sign-consistent (or equivalently, non-negative[13]) RIP matrix.

The only known non-negative RIP matrices that we are aware of are binary: all of their entries are from $\{0, 1\}$ after appropriately scaling [DeV07, Iwe09, BIS12]. Note that, in real-life applications, sparse and binary RIP matrices are sometimes desirable, see for instance [WHKI12]. The deterministic construction from the works [DeV07, Iwe09, BIS12] yield $m = \tilde{O}(\varepsilon^{-2} k^2 \log^2(d/k))$. Using the general coherence framework from [BIS12], simple argument shows that

**Fact 6.3.** *A random $m \times d$ binary matrix with $m = O(\varepsilon^{-2} k^2 \log(d/k))$ rows and $s = O(\varepsilon^{-1} k \log(d/k))$ ones per column satisfies (after normalization) the $(k, \varepsilon)$-RIP property with high probability.*[14]

When equipped with the Krahmer-Ward reduction above, such non-negative RIP matrices only give sign-consistent JL matrices with $m = O(\varepsilon^{-2} \log^2(1/\delta) \log d)$, worse than our Theorem 1.

---

[12]That is, letting $d$ be a random Rademacher sequence from $\{-1, 1\}^d$, then $A' = A \cdot \text{diag}(d)$.

[13]In the world of RIP matrices, *sign-consistent* constructions are equivalent to *non-negative* constructions, because we can arbitrarily flip the signs of the $k$-sparse vectors $x$ in order to convert a sign-consistent RIP matrix $A$ into a non-negative one $A'$ satisfying the same $(k, \varepsilon)$-RIP property.

[14]Note that the dependence on $k$ in $m = O(\varepsilon^{-2} k^2 \log(d/k))$ is quadratic, comparing to $m = O(\varepsilon^{-2} k \log(d/k))$ when the matrix can have negative entries. This "quadratic loss" is known as a singularity point for RIP matrices [AGR14].

**Lower Bound for Non-Negative RIP Matrices.** We show, in fact, there is a matching lower bound for all non-negative RIP matrices, including the binary ones. Thus, to construct an efficient sign-consistent JL matrix, one *must not* use the Krahmer-Ward reduction.

> **Theorem 4.** *There is some fixed $0 < \varepsilon_0 < 1/2$ so that the following holds. For any $0 < \varepsilon < \varepsilon_0$ and $m < O(\frac{d}{\varepsilon \log(k/\varepsilon)})$, let $A \in \mathbb{R}^{m \times d}$ be a non-negative $(k, \varepsilon)$-RIP matrix.*
> 
> *(a) At least half of the columns $A$ must have $\ell_1$-sparsity being $\Omega\left(\sqrt{\frac{k \log(d/k)}{\varepsilon}} / \log\left(m / \log(d/k)\right)\right)$.*
> 
> *(b) We must have $m = \Omega\left(\frac{k^2 \log(d/k)}{\varepsilon^2} / \log\left(\varepsilon^{-2} k / \log(d/k)\right)\right)$.*

We defer its proof to Appendix C and interpret the significance of Theorem 4 as follows.

- First, there is no known tight lower bounds on $m$ even for binary RIP matrices. Chandar has shown that $m = \Omega(\frac{k^2}{\varepsilon})$ [Cha08], while for general RIP matrices, we know $m = \Omega(\frac{k \log(d/k)}{\varepsilon})$ and $m = \Omega(\frac{k}{\varepsilon^2})$ from [NN13]. Theorem 4 states that the random binary RIP construction in Fact 6.3 is essentially optimal on $m$ among the class of non-negative RIP matrices.

- Second, there is no known tight lower bounds on the sparsity of non-negative RIP matrices, even in the binary case. The best known lower bound of $\Omega(k \log(d/k))$ is for the general case, and our sparsity lower bound $\tilde{\Omega}(\varepsilon^{-1} k \log(d/k))$ matches Fact 6.3.

- Third, one may ideally hope that Theorem 4 could be deduced in a black-box way from our Theorem 2 and the Krahmer-Ward reduction. Our result indicates that such a path must incur a log factor loss.

# Appendix

## A  Proof of Theorem 1: Upper Bound for Sign-Consistent JL Matrices

### A.1  Proof of Lemma 4.1

Racall that we want to upper bound

$$s^t \cdot \mathbb{E}[Z^t] = \sum_{\substack{i_1,\ldots,i_t,j_1,\ldots,j_t \in [d] \\ i_1 \neq j_1,\ldots,i_t \neq j_t}} \left(\prod_{u=1}^{t} x_{i_u} x_{j_u}\right) \left(\mathbb{E}_\sigma \prod_{u=1}^{t} \sigma_{i_u} \sigma_{j_u}\right) \left(\mathbb{E}_\eta \prod_{u=1}^{t} \sum_{r=1}^{m} \eta_{r,i_u} \eta_{r,j_u}\right), \qquad (A.1)$$

We first show that

**Lemma 4.1.**

$$s^t \cdot \mathbb{E}[Z^t] \leq e^t \sum_{v=2}^{t} \sum_{G \in \mathcal{G}''_{v,t}} \left(\frac{1}{t^t} \prod_{p=1}^{v} \sqrt{d_p}^{d_p}\right) \cdot \sum_{r_1,\ldots,r_t \in [m]} \prod_{i=1}^{w} \left(\frac{s}{m}\right)^{v_i} .$$

*Here,*

- $\mathcal{G}''_{v,t}$ *is a set of directed multigraphs with $v$ labeled vertices (1 to $v$) and $t$ labeled edges (1 to $t$).*
- $d_p$ *is the total degree of vertex $p \in [v]$ in a graph $G \in \mathcal{G}''_{v,t}$.*[15]
- $w$ *and $v_1, \ldots, v_w$ are defined by $G$ and $r_1, \ldots, r_t$ as follows. Let an edge $u \in [t]$ be colored with $r_u \in [m]$, then we define $w$ to be the number of distinct colors used in $r_1, \ldots, r_t$, and $v_i$ to be the number of vertices incident to an edge with color $i \in [w]$.*

*Proof.* We prove the desired inequality from (A.1) in three steps. The first step removes the random variables of $\sigma$ in (A.1). The second step removes $x$ from (A.1) using the assumption of $\|x\|_2 = 1$. The third step removes the random variables $\eta$ in (A.1) by carefully exploiting the independence or negative correlation among different $\eta$ terms.

**In the first step**, we use a standard trick to map each summand

$$\left(\prod_{u=1}^{t} x_{i_u} x_{j_u}\right) \left(\mathbb{E}_\sigma \prod_{u=1}^{t} \sigma_{i_u} \sigma_{j_u}\right) \left(\mathbb{E}_\eta \prod_{u=1}^{t} \sum_{r=1}^{m} \eta_{r,i_u} \eta_{r,j_u}\right)$$

in expression (A.1) to a directed multigraph. That is, for each pair of $(i_u, j_u)$ where $u \in [t]$, we associate it with a directed edge $i_u \to j_u$. It is easy to see that it suffices for us to consider only graphs with all the vertices having *even* total degree, since otherwise the expectation $\mathbb{E}_\sigma \prod_{u=1}^{t} \sigma_{i_u} \sigma_{j_u}$ becomes zero (e.g., $\mathbb{E}_\sigma[\sigma_1^3 \sigma_2^2 \sigma_4] = 0$).

To make this precise, let us define $\mathcal{G}_t$ to be the set of directed multigraphs $G$ with the following properties:

- $G$ has between 2 and $t$ (identical) vertices.
- $G$ has exactly $t$ *distinct* edges, labels by $1, 2, \ldots, t$.
- There are no self-loops.

---

[15]The total degree of a vertex is defined as the number of incident edges regardless of direction.

- Each vertex has a non-zero and even total degree (sum of in- and out-degrees).

Note that we intentionally made the vertices *identical (i.e., unlabeled)* in the above definition, and we will separately enumerate over the vertex labeling.

Let $f$ be a map from $(i_u, j_u)_{u \in [t]}$ to its underlying graph $G$ by adding a directed edge $i_u \to j_u$ as the $u$-th edge of a graph. Our argument above shows that in order to enumerate $(i_u, j_u)_{u \in [t]}$ in (A.1), it suffices to enumerate $G \in \mathcal{G}_t$ and the vertex labeling as follows

$$s^t \cdot \mathbb{E}[Z^t] = \sum_{\substack{G \in \mathcal{G}_t \\ f((i_u,j_u)_{u=1}^t)=G}} \sum_{i_1 \neq j_1, \ldots, i_t \neq j_t \in [d]} \left( \prod_{u=1}^t x_{i_u} x_{j_u} \right) \left( \mathbb{E}_\eta \prod_{u=1}^t \sum_{r=1}^m \eta_{r,i_u} \eta_{r,j_u} \right)$$

In the above expression, the $\mathbb{E}_\sigma \prod_{u=1}^t \sigma_{i_u} \sigma_{j_u}$ factors have disappeared because they equal to one if $G$ has even total degrees for all of its vertices. Also, the second summation —the one over all choices of $(i_u, j_u)_u$ such that $f((i_u, j_u)) = G$— is in fact an enumeration over the missing vertex labeling of the graph $G$.

**In the second step**, we observe that $\eta_{\star,i}$ and $\eta_{\star,j}$ for $i \neq j$ are independent because they are for different columns, and generated by the same random process. Thus, for a given graph $G \in \mathcal{G}_t$, the $\mathbb{E}_\eta \prod_{u=1}^t \sum_{r=1}^m \eta_{r,i_u} \eta_{r,j_u}$ factor has the same value for all mappings with $f((i_u, j_u)_{u=1}^t) = G$ (i.e., for all the vertex labeling).[16] Let us call this function $\hat{\eta}(G)$ and write:

$$s^t \cdot \mathbb{E}[Z^t] = \sum_{\substack{G \in \mathcal{G}_t \\ f((i_u,j_u)_{u=1}^t)=G}} \sum_{i_1 \neq j_1, \ldots, i_t \neq j_t \in [d]} \left( \prod_{u=1}^t x_{i_u} x_{j_u} \right) \hat{\eta}(G) = \sum_{G \in \mathcal{G}_t} \hat{\eta}(G) \cdot \sum_{\substack{i_1 \neq j_1, \ldots, i_t \neq j_t \in [d] \\ f((i_u,j_u)_{u=1}^t)=G}} \left( \prod_{u=1}^t x_{i_u} x_{j_u} \right) .$$

(A.2)

Next, for a fixed graph $G \in \mathcal{G}_t$, let $v$ be the number of vertices in $G$ and $d_p$ the total degree of vertex $p \in [v]$. We observe a simple fact that

$$\binom{t}{d_1/2, \ldots, d_v/2} \cdot \sum_{\substack{i_1 \neq j_1, \ldots, i_t \neq j_t \in [d] \\ f((i_u,j_u)_{u=1}^t)=G}} \left( \prod_{u=1}^t x_{i_u} x_{j_u} \right) \leq \left( \sum_{l=1}^d x_l^2 \right)^t \cdot v! = v! . \quad (A.3)$$

The above inequality holds as each (distinct) monomial in $\sum_{i_1 \neq j_1, \ldots, i_t \neq j_t \in [d], f((i_u,j_u)_{u=1}^t)=G} \left( \prod_{u=1}^t x_{i_u} x_{j_u} \right)$, for instance appears at most $v!$ times in this summation due to vertex re-labeling, and thus $\binom{t}{d_1/2, \ldots, d_v/2} \cdot v!$ times in total on the left hand side; each of these monomials also appear on the right hand side exactly $\binom{t}{d_1/2, \ldots, d_v/2} \cdot v!$ times; and finally, each monomial is non-negative and $\|x\|_2 = 1$.

Now we are ready to plug (A.3) to (A.2) and get

$$s^t \cdot \mathbb{E}[Z^t] = \sum_{\substack{G \in \mathcal{G}_t \\ f((i_u,j_u)_{u=1}^t)=G}} \sum_{i_1 \neq j_1, \ldots, i_t \neq j_t \in [d]} \left( \prod_{u=1}^t x_{i_u} x_{j_u} \right) \hat{\eta}(G) \leq \sum_{G \in \mathcal{G}_t} \frac{v!}{\binom{t}{d_1/2, \ldots, d_v/2}} \hat{\eta}(G)$$

$$= \sum_{G \in \mathcal{G}_t'} \frac{1}{\binom{t}{d_1/2, \ldots, d_v/2}} \hat{\eta}(G) \quad (A.4)$$

$$\leq e^t \sum_{G \in \mathcal{G}_t'} \frac{1}{t^t} \prod_{p=1}^v \sqrt{d_p}^{d_p} \hat{\eta}(G) \quad (A.5)$$

---

[16] In fact, the value does not depend on the edge labeling of $G$ as well, but we are not going to use this fact.

Here in (A.4), we have defined $\mathcal{G}'_t$ to be the same as $\mathcal{G}_t$ except that we require the $v$ vertices to have distinct labels in $[v]$, and (A.4) follows because each there are $v!$ distinct ways to label each $G \in \mathcal{G}_t$.[17] For (A.5), we use that $t! \geq t^t/e^t$ and $\prod_{p=1}^{v}(d_p/2)! \leq \prod_{p=1}^{v} \sqrt{d_p}^{d_p}$. We have been ambiguous when writing $\hat{\eta}(G)$ because $G$ may either be vertex-labelled or not vertex-labelled; its value is independent of such a labeling.

**In the third step**, we give an upper bound on $\hat{\eta}(G)$ by carefully exploiting the independence or negative correlation among the random variables in it. We first rewrite

$$\hat{\eta}(G) = \mathbb{E}_\eta \prod_{u=1}^{t} \sum_{r=1}^{m} \eta_{r,i_u}\eta_{r,j_u} = \sum_{r_1,\ldots,r_t \in [m]} \mathbb{E}_\eta \prod_{u=1}^{t} \eta_{r_u,i_u}\eta_{r_u,j_u}$$

From this point, whenever we fix a graph $G$ and a sequence $r = (r_1, \ldots, r_t) \in [m]^t$, we would like to view them together as a *directed and edge-colored multigraph* $(G, r)$ —i.e., graph $G$ appended with edge colors such that its $u$-th edge $i_u \to j_u$ is given the color $r_u \in [m]$.

The big advantage of such edge coloring is to allow us to exploit the negative correlation between graphs of different colors. Indeed, for any fixed $G \in \mathcal{G}_t$ and $r \in [m]^t$, let us define

$$\tilde{\eta}_c(G, r) \stackrel{\text{def}}{=} \prod_{u \in [t], r_u = c} \eta_{r_u,i_u}\eta_{r_u,j_u}$$

to be the factors of $\eta$ associated with color $c \in [m]$. Then we have

$$\hat{\eta}(G) = \sum_{r_1,\ldots,r_t \in [m]} \mathbb{E}_\eta \prod_{c=1}^{m} \tilde{\eta}_c(G, r) \leq \sum_{r_1,\ldots,r_t \in [m]} \prod_{c=1}^{m} \mathbb{E}_\eta[\tilde{\eta}_c(G, r)]$$

Here the inequality is owing to the fact that different rows of $\eta$ are negatively correlated.[18]

Next, let us denote by $w \in [t]$ the number of distinct colors in $(G, r)$. For notational simplicity, we can assume that the used colors in $G$ are $1, 2, \ldots, w$ (so $w+1, \ldots, m$ are unused). Let $G_i$ be the subgraph of $G$ containing all the edges of color $i \in [w]$, and suppose that $G_i$ has $v_i \geq 2$ vertices and $c_i \geq 1$ edges.

It is straightforward to see that for a fixed color $i \in [w]$, there are precisely $v_i$ distinct $\eta$ factors in the definition of $\tilde{\eta}_i(G, r)$ (by the definition that $G_i$ has $v_i$ "vertices"). Since these $\eta$ factors are across different columns, they are independent and each has a probability of $\frac{s}{m}$ to be 1 (due to our probabilistic construction of $\mathcal{A}$). We therefore can simply write $\mathbb{E}_\eta[\tilde{\eta}_i(G, r)] = \left(\frac{s}{m}\right)^{v_i}$ and conclude that

$$\hat{\eta}(G) \leq \sum_{r_1,\ldots,r_t \in [m]} \prod_{i=1}^{w} \left(\frac{s}{m}\right)^{v_i} \tag{A.6}$$

---

[17] All these labelings are distinct in $\mathcal{G}'_t$ because there is a canonical way to label the vertices of each $G \in \mathcal{G}_t$: since $G$ does not have isolated vertices, to get a canonical labeling, we can order the directed edges in $G$ in increasing order and label the vertices in this order as well.

[18] As a simple example, we have $\mathbb{E}[\eta_{2,4} \cdot \eta_{2,5} \cdot \eta_{3,4} \cdot \eta_{3,7}] \leq \mathbb{E}[\eta_{2,4} \cdot \eta_{2,5}] \cdot \mathbb{E}[\eta_{3,4} \cdot \eta_{3,7}]$ because: (a) $\eta_{2,4}$ is negatively correlated with $\eta_{3,4}$, and independent with $\eta_{3,7}$, and (b) $\eta_{2,5}$ is independent with both $\eta_{3,4}$ and $\eta_{3,7}$. In general, if an indicator variable is set to 1, the probability of other indicator variables being set to 1 in the same column and different row, decreases. Therefore, the product of expectations is always no less than the expectation of product of corresponding negatively correlated terms.

At last, we incorporate (A.6) in (A.5) and get

$$s^t \cdot \mathbb{E}[Z^t] \leq e^t \sum_{G \in \mathcal{G}'_t} \left( \frac{1}{t^t} \prod_{p=1}^{v} \sqrt{d_p}^{d_p} \right) \cdot \sum_{r_1,\ldots,r_t \in [m]} \prod_{i=1}^{w} \left( \frac{s}{m} \right)^{v_i}$$

$$\leq e^t \sum_{v=2}^{t} \sum_{G \in \mathcal{G}''_{v,t}} \left( \frac{1}{t^t} \prod_{p=1}^{v} \sqrt{d_p}^{d_p} \right) \cdot \sum_{r_1,\ldots,r_t \in [m]} \prod_{i=1}^{w} \left( \frac{s}{m} \right)^{v_i} \quad . \tag{A.7}$$

Here $\mathcal{G}''_{v,t}$ contains graphs with $v$ labeled vertices and $t$ labeled edges, without the restriction (like we did in $\mathcal{G}_t$ and $\mathcal{G}'_t$) that a vertex has a positive or even degree. We can have $v \leq t$ because in $\mathcal{G}'_t$ each vertex must degree no less than 2, while the total degree over all vertices equal to $2t$. Therefore, going from $\mathcal{G}'$ to $\mathcal{G}''$ we only add non-negative terms and the inequality goes through. This concludes the proof of Lemma 4.1. □

## A.2 Proof of Lemma 4.2

Recall that in Section 4 we proceed from Lemma 4.1 as follows. Instead of enumerating $G \in \mathcal{G}''_{v,t}$ as a whole, we now enumerate subgraphs of different colors separately, and then combine the results. Below is one way (and perhaps the only way the authors believe without incurring a $\log(1/\delta)$ factor loss in $m$) to enumerate $G$ that can lead to tight upper bounds

$$s^t \cdot \mathbb{E}[Z^t] \leq e^t \underbrace{\sum_{v=2}^{t}}_{\text{i}} \underbrace{\sum_{w=1}^{t} \binom{m}{w}}_{\text{ii}} \underbrace{\sum_{\substack{c_1,\ldots,c_w \\ c_1+\ldots+c_w=t \\ c_i \geq 1}} \binom{t}{c_1,\ldots,c_w}}_{\text{iii}} \underbrace{\sum_{\substack{v_1,\ldots,v_w \\ 2 \leq v_i \leq 2c_i}} \left( \frac{s}{m} \right)^{v_1+\cdots+v_w}}_{\text{iv}} \underbrace{\sum_{f_1,\ldots,f_w}}_{\text{v}} \underbrace{\forall i, G_i \in \mathcal{G}''_{v_i,c_i}}_{\text{vi}} \frac{1}{t^t} \prod_{p=1}^{v} \sqrt{d_p}^{d_p} \tag{4.1}$$

This gigantic expression enumerates all graphs $G \in \mathcal{G}''_{v,t}$ and its coloring $r_1,\ldots,r_t \in [m]$ in six steps:

(i). Number of graph vertices, $v \in \{2,\ldots,t\}$; the vetices are labelled by $1,2,\ldots,v$.

(ii). Number of used edge colors, $w \in \{1,\ldots,t\}$, and all $\binom{m}{w}$ possibilities of choosing $w$ colors.

(iii). Edge colorings of the graph using selected $w$ colors: how many (denoted by $c_i \geq 1$) edges are colored in color $i$ and which of the $t$ edges are colored in color $i$.

(iv). Number of vertices $v_i \in \{2,\ldots,2c_i\}$ in each $G_i$, the subgraph containing edges of color $i$.

(v). All possible increasing functions $f_i : [v_i] \to [v]$, such that $f_i(j)$ maps vertex $j$ in $G_i$ to the $f_i(j)$-th global vertex. (And we ensure $f_i(j) < f_i(k)$ for $j < k$ to reduce double counting.)

(vi). All graphs $G_i \in \mathcal{G}''_{v_i,c_i}$ with $v_i$ labeled vertices (1 to $v_i$) and $c_i$ labeled edges (1 to $c_i$).

(Using all the information above, $d_p$, the degree of vertex $p \in [v]$ is well defined.)

We emphasize here that any pair of graph $G \in \mathcal{G}''_{v,t}$ and coloring $r_1,\ldots,r_t \in [m]$ will be generated *at least once* in the above procedure.[19] Thus, (4.1) follows from Lemma 4.1, since the summation terms also have the same value $\left(\frac{s}{m}\right)^{v_1+\cdots+v_w} \frac{1}{t^t} \prod_{p=1}^{v} \sqrt{d_p}^{d_p}$.

It is now possible to consider $G_i$'s separately in (4.1) and prove the following lemma:

---

[19]This follows from the fact that $G$ and $r_1,\ldots,r_t$ together determine (a) $w$, the number of used colors, (b) $G_i$ for each $i \in [w]$ (with $v_i$ vertices and $c_i$ edges), the subgraph of $G$ of the $i$-th used color, and (c) $f_i$, the vertex mapping from $G_i$ back to $G$. Any such triple will be generated least once in (4.1). Note also, we may have double counts but it will not affect our asymptotic upper bound.

**Lemma 4.2.** *From (4.1) we can get*

$$s^t \cdot \mathbb{E}[Z^t] \leq 2^{O(t)} \sum_{v=2}^{t} \sum_{w=1}^{t} \binom{m}{w} \sum_{\substack{c_1,\ldots,c_w \\ c_1+\ldots+c_w=t \\ c_i \geq 1}} \binom{t}{c_1,\ldots,c_w} \sum_{\substack{v_1,\ldots,v_w \\ 2 \leq v_i \leq 2c_i}} \prod_{j=1}^{w} \left(\frac{s}{m}\right)^{v_j} v_j^{c_j} \binom{v-1}{v_j-1}$$

*Proof.* From (4.1) it suffices to show that

$$\sum_{f_1,\ldots,f_w} \sum_{\forall i, G_i \in \mathcal{G}''_{v_i,c_i}} \frac{1}{t^t} \prod_{p=1}^{v} \sqrt{d_p}^{d_p} \leq 2^{O(t)} \cdot \prod_{j=1}^{w} v_j^{c_j} \binom{v-1}{v_j-1} \qquad (A.8)$$

Recall that here $d_p$ remains to be the total degree of vertex $p \in [v]$ in the combined graph $G$, which is essentially $G_1 \cup \cdots \cup G_w$ but glued together using the vertex mappings $f_1,\ldots,f_w$.

To show (A.8), let us define:

$$\text{for any } \vec{\gamma} \in \mathbb{Z}_{\geq 0}^w \text{ and } \vec{a} \in \mathbb{R}_{>0}^v: \qquad S(\vec{\gamma},\vec{a}) \stackrel{\text{def}}{=} \sum_{\forall i, G_i \in \mathcal{G}''_{v_i,\gamma_i}} \prod_{p=1}^{v} \sqrt{a_p}^{d_p} \;,$$

where as before $d_p$ is the degree of the $p$-th vertex in the combined graph $G = G_1 \cup \cdots \cup G_w$, but $a_p$ is a constant. Ideally, we want an upper bound on $S(\vec{\gamma},\vec{a})$ for the choice of $\vec{\gamma} = \vec{c}$ and $\vec{a} = \vec{d}$, so that $S(\vec{\gamma},\vec{a})$ becomes identical to the left hand side of (A.8).[20] Thus, let us now shoot for an upper bound of $S(\vec{\gamma},\vec{a})$ using induction on $\vec{\gamma}$.

When $\vec{\gamma} = \vec{0}$, observe that $S(\vec{0},\vec{a}) = 1$ since each $G_i$ has no edge in it and $d_p = 0$ for all $p \in [v]$.

Now, consider adding an edge to $G$ of some color $l$. for any $\vec{\gamma}$, define $\vec{\gamma}'$ so that $\gamma'_l = \gamma_l + 1$ and $\forall j \neq l : \gamma'_j = \gamma_j$. Then,

$$\frac{S(\vec{\gamma}',\vec{a})}{S(\vec{\gamma},\vec{a})} \leq \sum_{\alpha \neq \beta \in [v_l]} \sqrt{a_{f_l(\alpha)}} \sqrt{a_{f_l(\beta)}} \leq \sum_{\alpha=1}^{v_l} \left(\sqrt{a_{f_l(\alpha)}}\right)^2 \leq \sum_{\alpha=1}^{v_l} \left(a_{f_l(\alpha)}\right) \cdot v_l$$

where the first inequality is because this new edge may be added anywhere between two vertices $f_l(\alpha)$ and $f_l(\beta)$ for $\alpha, \beta \in [v_l]$, the second inequality is by the simple expansion of square of sum, the last inequality is by Cauchy-Schwartz. Therefore, by induction we conclude that

$$\sum_{\forall i, G_i \in \mathcal{G}''_{v_i,c_i}} \prod_{p=1}^{v} \sqrt{a_p}^{d_p} = S(\vec{c},\vec{a}) \leq \prod_{j=1}^{w} \left(\sum_{\alpha \in [v_j]} a_{f_j(\alpha)}\right)^{c_j} \cdot v_j^{c_j} \;. \qquad (A.9)$$

It is worth noting that (A.9) would be sufficient for us to show (A.8), if one could replace $\vec{a}$ by $\vec{d}$. However, since the degree vector $\vec{d}$ is determined *after* the choices of $G_j$ for $j \in [w]$, this simple substitution is impossible and we need a different approach.

Indeed, we fix this by enumerating $G_i \in \mathcal{G}''_{v_i,c_i}$ in two steps: first enumerating the degrees $d'_1,\ldots,d'_v$ and then enumerating the possible $G_i$'s satisfying such degree spectrum (i.e., $d_p = d'_p$ for all $p \in [v]$)

$$\sum_{\forall i, G_i \in \mathcal{G}''_{v_i,c_i}} \prod_{p=1}^{v} \sqrt{d_p}^{d_p} = \sum_{\substack{d'_1,\ldots,d'_v \geq 0 \\ d'_1+\cdots+d'_v=2t}} \left( \sum_{\substack{\forall i, G_i \in \mathcal{G}''_{v_i,c_i} \\ \text{s.t.} \forall p, d_p=d'_p}} \prod_{p=1}^{v} \sqrt{d'_p}^{d_p} \right)$$

---
[20] However, this mission is non-trivial because the values of $\vec{d}$ are decided *after* $G_i \in \mathcal{G}''_{v_i,c_i}$ are chosen. Let us anyways ignore this issue for a moment and resolve it later.

This seemingly redundant separation in fact enables us to prove (A.8). Indeed, we proceed the above equation as follows

$$\sum_{\forall i, G_i \in \mathcal{G}''_{v_i, c_i}} \prod_{p=1}^{v} \sqrt{d_p}^{d_p} \leq \sum_{\substack{d'_1, \ldots, d'_v \geq 0 \\ d'_1 + \cdots + d'_v = 2t}} \left( \sum_{\forall i, G_i \in \mathcal{G}''_{v_i, c_i}} \prod_{p=1}^{v} \sqrt{d'_p}^{d_p} \right)$$

$$= \sum_{\substack{d'_1, \ldots, d'_v \geq 0 \\ d'_1 + \cdots + d'_v = 2t}} S(\vec{c}, \vec{d'}) \leq \sum_{\substack{d'_1, \ldots, d'_v \geq 0 \\ d'_1 + \cdots + d'_v = 2t}} \prod_{j=1}^{w} \left( \sum_{\alpha \in [v_j]} d'_{f_j(\alpha)} \right)^{c_j} \cdot v_j^{c_j} . \quad \text{(A.10)}$$

Here the first inequality gets rid of the $d_p = d'_p$ constraint, and the second one is from (A.9).

To proceed from here, we make use of the summation over $f_1, \ldots, f_w$ that we intentionally ignored when defining $S(\vec{\gamma}, \vec{a})$, and get

$$\sum_{f_1, \ldots, f_w} \prod_{j=1}^{w} \left( \sum_{\alpha \in [v_j]} d'_{f_j(\alpha)} \right)^{c_j} \cdot v_j^{c_j} = \prod_{j=1}^{w} v_j^{c_j} \sum_{f_j} \left( \sum_{\alpha \in [v_j]} d'_{f_j(\alpha)} \right)^{c_j}$$

$$\leq \prod_{j=1}^{w} v_j^{c_j} \binom{v-1}{v_j - 1} \cdot (2t)^{c_j} = (2t)^t \prod_{j=1}^{w} v_j^{c_j} \binom{v-1}{v_j - 1} \quad \text{(A.11)}$$

Above, the first equality is a simple swap between adacant $\sum$ and $\prod$. The inequality in (A.11) needs some justifications:

Recall that the mapping $f_j$ chooses $v_j$ vertex labels out of $[v]$. If we represent $2t$ as the summation $d'_1 + \cdots + d'_v$, we have that $\sum_{\alpha \in [v_j]} d'_{f_j(\alpha)}$ is the partial sum over only the selected $v_j$ vertices under $f_j$. Hence, for a fixed $f_j$, each monomial in the expansion of $\left( \sum_{\alpha \in [v_j]} d'_{f_j(\alpha)} \right)^{c_j}$ also appears in $(2t)^{c_j} = (d'_1 + \cdots + d'_v)^{c_j}$ with the same coefficient. However, any such monomial can appear in at most $\binom{v-1}{v_j - 1}$ different $f_j$ mappings: each such monimal contains at least one vertex (so may look like $(d'_p)^{c_j}$ for some $p \in [v]$), and $f_j$ could have the freedom to pick at most the $v_j - 1$ more vertices out of $v - 1$ to complete as an increasing mapping $[v_j] \to [v]$.

Finally, we plug (A.11) into (A.10) and get

$$\sum_{f_1, \ldots, f_w} \sum_{\forall i, G_i \in \mathcal{G}''_{v_i, c_i}} \frac{1}{t^t} \prod_{p=1}^{v} \sqrt{d_p}^{d_p} \leq \sum_{\substack{d'_1, \ldots, d'_v \geq 0 \\ d'_1 + \cdots + d'_v = 2t}} \frac{(2t)^t}{t^t} \prod_{j=1}^{w} v_j^{c_j} \binom{v-1}{v_j - 1} \leq 2^{O(t)} \cdot \prod_{j=1}^{w} v_j^{c_j} \binom{v-1}{v_j - 1}$$

where the last inequality is because the number of ways to partition $2t$ into $d'_1 + \cdots + d'_v$ less than $2^{O(t+v)} = 2^{O(t)}$. This concludes (A.8) and thus the proof of Lemma 4.2. □

## A.3 Proof of Lemma 4.3

The last lemma of our proof is essentially to handle algebra manipulations in a careful way.

**Lemma 4.3.** *We can rearrange the inequality in (4.2) and get*

$$s^t \cdot \mathbb{E}[Z^t] \leq 2^{O(t)} \cdot t^t \left( \frac{s^2}{m} \right)^t .$$

*Proof.* Simplifying the result of Lemma 4.2, we get:

$$s^t \cdot \mathbb{E}[Z^t] \leq 2^{O(t)} \sum_{v=2}^{t} \sum_{w=1}^{t} \binom{m}{w} \sum_{\substack{c_1,\ldots,c_w \\ c_1+\ldots+c_w=t \\ c_i \geq 1}} \binom{t}{c_1,\ldots,c_w} \sum_{\substack{v_1,\ldots,v_w \\ 2 \leq v_i \leq 2c_i}} \prod_{j=1}^{w} \left(\frac{s}{m}\right)^{v_j} v_j^{c_j} v^{v_j-1} \qquad (A.12)$$

$$= 2^{O(t)} \sum_{v=2}^{t} \sum_{w=1}^{t} \binom{m}{w} \sum_{\substack{c_1,\ldots,c_w \\ c_1+\ldots+c_w=t \\ c_i \geq 1}} \binom{t}{c_1,\ldots,c_w} \prod_{j=1}^{w} \frac{1}{v} \sum_{v_j=2}^{2c_j} \left(\frac{sv}{m}\right)^{v_j} v_j^{c_j}$$

$$\leq 2^{O(t)} \sum_{v=2}^{t} \sum_{w=1}^{t} \binom{m}{w} \sum_{\substack{c_1,\ldots,c_w \\ c_1+\ldots+c_w=t \\ c_i \geq 1}} \binom{t}{c_1,\ldots,c_w} \prod_{j=1}^{w} \frac{1}{v} \left(\frac{sv}{m}\right)^2 (2c_j)^{c_j+1} \qquad (A.13)$$

$$\leq 2^{O(t)} \sum_{v=2}^{t} \sum_{w=1}^{t} \binom{m}{w} \sum_{\substack{c_1,\ldots,c_w \\ c_1+\ldots+c_w=t \\ c_i \geq 1}} \frac{t^t}{c_1^{c_1} \cdot \ldots \cdot c_w^{c_w}} \prod_{j=1}^{w} \left(\frac{s^2 v}{m^2}\right) c_j^{c_j+1} \qquad (A.14)$$

$$= 2^{O(t)} \sum_{v=2}^{t} \sum_{w=1}^{t} \binom{m}{w} \sum_{\substack{c_1,\ldots,c_w \\ c_1+\ldots+c_w=t \\ c_i \geq 1}} t^t \left(\frac{s^2 v}{m^2}\right)^w \prod_{j=1}^{w} c_j \qquad (A.15)$$

Here, (A.12) uses the upper bound on binomial coefficients. To get (A.13), we require $st < m$.[21] Then, since $v \leq t$, it satisfies that $\frac{sv}{m} < 1$ and we can replace the power on $\left(\frac{sv}{m}\right)^{v_j}$ by 2, to get an upper bound $\left(\frac{sv}{m}\right)^2$. To obtain (A.14), we use Stirling's formula to bound the factorials in $\binom{t}{c_1,\ldots,c_w}$, and $2^{c_1+\cdots+c_w+w} = 2^{O(t)}$.

The multiplicant $\prod_{j=1}^{w} c_j$ in (A.15) can be upper bounded by $\left(\frac{t}{w}\right)^w$, since $c_1 + \cdots + c_w = t$. Also, the number of choices of positive integers $c_1, \ldots, c_w$ summing up to $t$ is $\binom{t-1}{w-1}$, upper bounded by $2^{O(w)} \left(\frac{t}{w}\right)^w \leq 2^{O(t)} \left(\frac{t}{w}\right)^w$. Incorporating these in (A.15) gives:

$$s^t \cdot \mathbb{E}[Z^t] \leq 2^{O(t)} \sum_{v=2}^{t} \sum_{w=1}^{t} \binom{m}{w} t^t \left(\frac{s^2 v}{m^2}\right)^w \left(\frac{t}{w}\right)^{2w}$$

$$\leq 2^{O(t)} \sum_{v=2}^{t} \sum_{w=1}^{t} t^t \left(\frac{s^2}{m}\right)^w \left(\frac{t}{w}\right)^{3w} \qquad (A.16)$$

$$\leq 2^{O(t)} \cdot t^t \left(\frac{s^2}{m}\right)^t \qquad (A.17)$$

Here to get (A.16), we again use the upper bound on binomial coefficients for $\binom{m}{w}$. For (A.17), note that $\left(\frac{t}{w}\right)^{3w}$ is maximized when $w = t/e$ (which can be seen by taking the derivative), so is upper bounded by $e^{3t/e} = 2^{O(t)}$. Therefore, we can replace $\left(\frac{s^2}{m}\right)^w$ by $\left(\frac{s^2}{m}\right)^t$ since this is at this moment the only term that depends on $w$. This concludes the proof of Lemma 4.3. □

---

[21]For our setting of parameters to be chosen later, this will correspond to $\varepsilon^{-1} \cdot 2C > 1$ for a large constant $C > 1$.

# B  Proof of Theorem 3: $\ell_1$-Sparsity Lower Bound for JL Matrices

Our proof to Theorem 3 follows from the same proof framework as [NN13]; however, since the $\ell_1$-sparsity guarantee is a stronger one, this strengthening needs a lot of careful care.

Given a matrix $A \in \mathbb{R}^{m \times n}$, let us denote its columns by $v_1, \ldots, v_n \in \mathbb{R}^m$. Throughout this section, we assume that $A$ is $\varepsilon$-incoherent: for all $i \neq j$, $|\langle v_i, v_j \rangle| \leq \varepsilon$, and for all $i$, $\|v_i\|_2 = 1$.

Like in the sparsity case from [NN13], we first need a weaker lower bound on the $\ell_1$ norm:

**Lemma B.1.** *Suppose $m < n/(40 \log(1/2\varepsilon))$, and $A \in \mathbb{R}^{m \times n}$ is $\varepsilon$-incoherent. If $A$ has $n/2$ columns with $\ell_1$ norm at most $\sqrt{s/2}$ each, then $s \geq 1/(4\varepsilon)$.*

*Proof.* For the sake of contradiction, assume $s < 1/(4\varepsilon)$.

Let $W \subset [n]$ be a set of the $n/2$ columns each with $\ell_1$ norm at most $\sqrt{s/2}$. Define $L(W) = \{(i,j) : A_{i,j}^2 \geq 2\varepsilon, j \in W\}$ to be the set of large entries, and $S(W) = \{(i,j) : A_{i,j}^2 < 2\varepsilon, j \in W\}$ to be that of small entries. Clearly, $S(W)$ and $L(W)$ are disjoint and span all the entries in $W$. Let us bound the sum of squares of matrix entries in $S(W)$ and $L(W)$ respectively.

- Small entries in $W$ are less than $\sqrt{2\varepsilon}$ in absolute magnitude, so their squares sum up to at most $\sqrt{2\varepsilon}$ times the $\ell_1$ norm of such entries:

$$\sum_{(i,j) \in S(W)} A_{i,j}^2 \leq \sqrt{2\varepsilon} \cdot \Big( \sum_{(i,j) \in S(W)} |A_{i,j}| \Big) = \sqrt{2\varepsilon} \cdot \Big( \sum_{i \in [m], j \in W} |A_{i,j}| - \sum_{(i,j) \in L(W)} |A_{i,j}| \Big)$$

$$\leq \sqrt{2\varepsilon} \cdot \Big( \sum_{j \in W} \|v_j\|_1 - \sqrt{2\varepsilon} |L(W)| \Big) \leq \frac{n}{2}\sqrt{\varepsilon s} - 2\varepsilon |L(W)| < \frac{n}{4} - 2\varepsilon |L(W)| \quad \text{(B.1)}$$

- For large entries, we reuse the same analysis as [NN13]. Let $X$ denote the square of a random entry from $L(W)$. Then,

$$\sum_{(i,j) \in L(W)} A_{i,j}^2 = |L(W)| \cdot \mathbb{E}[X] = |L(W)| \cdot \int_{x=0}^{1} \Pr[X > x] dx \leq 2\varepsilon|L(W)| + m \cdot \int_{x=2\varepsilon}^{1} \frac{10}{x} dx$$

$$= 2\varepsilon|L(W)| + 10m \log(1/2\varepsilon) \leq 2\varepsilon|L(W)| + \frac{n}{4} \quad . \quad \text{(B.2)}$$

Here the first inequality is due to a simple fact about $\varepsilon$-incoherent matrices: there cannot be more than $\frac{10}{x}$ entries in each row of absolute value more than $\sqrt{x}$, for any $x \geq 2\varepsilon$. See for instance [NN13, Lemma 3]. The second inequality is owing to the choice of $m < n/(40 \log(1/2\varepsilon))$.

Now we can combine equations (B.1) and (B.2) to get $\sum_{(i,j) \in S(W) \cup L(W)} A_{i,j}^2 < \frac{n}{2}$. On the other hand, $\sum_{(i,j) \in S(W) \cup L(W)} A_{i,j}^2 = \sum_{j \in W} \|v_j\|_2^2 = \frac{n}{2}$ and we get a contradiction. □

The above lower bound is weak since it is obtained merely from a counting argument. Let us now strengthen it into a stronger form, by using the pigeon-hole principle to find $N$ columns that pairwisely and positively correlate to each other. This cannot happen if the original matrix is $\varepsilon$-incoherent. Let us explain:

**Lemma B.2.** *Let $0 < \varepsilon < 1/2$, $A \in \mathbb{R}^{m \times n}$ be an $\varepsilon$-incoherent matrix, and $s$ be any value such that half of $A$'s columns have $\ell_1$ norm at most $\sqrt{s}/2$. Define $C = 2/(1 - 1/\sqrt{2})$. Then, for any $t \in [s/2]$ with $t/s > C\varepsilon$, we must have $s \geq t(N-1)/C$ with $N = \left\lceil \frac{n}{2^{t+1} \binom{m}{t} \binom{2(s+t)}{t}} \right\rceil$.*

*Proof.* The proof structure is similar to that of [NN13, Lemma 9]: for any vector $v_i$, consider its $t$ largest coordinates in absolute magnitude, and define its *t-type* to be a triple containing:

- the locations of the top $t$ coordinates (there are at most $\binom{m}{t}$ choices);
- the signs of the top $t$ coordinates (there are at most $2^t$ different choices); and
- the rounding of the top $t$ values so that their squares round to the nearest integer multiple of $1/(2s)$. Values halfway between two multiples can be rounded arbitrarily. (There are at most $\binom{2s+2t}{t}$ number of different roundings.[22])

All in all, there are $2^t \binom{m}{t}\binom{2s+2t}{t}$ possible *t-types*. By the pigeon-hole principle, out of $n/2$ column vectors that have $\ell_1$ norm at most $\sqrt{s}/2$, we can select $N$ vectors $\tilde{v}_1, \ldots, \tilde{v}_N$, such that they all have the same *t-type*.

Let $S \subset [n]$ be the set of the largest coordinates for these vectors, and we have $|S| = t$. Now define $u_i = (\tilde{v}_i)_{[m]-S} \in \mathbb{R}^{m-t}$, with the coordinates in $S$ zeroed out. Then, for $j \neq k \in [N]$, since $\tilde{v}_j$ and $\tilde{v}_k$ have the same type, we must have

$$\langle u_j, u_k \rangle = \langle \tilde{v}_j, \tilde{v}_k \rangle - \sum_{r \in S} (\tilde{v}_j)_r (\tilde{v}_k)_r \leq \varepsilon - \sum_{r \in S} (\tilde{v}_j)_r ((\tilde{v}_j)_r \pm 1/\sqrt{2s})$$

$$\leq \varepsilon - \sum_{r \in S} \left((\tilde{v}_j)_r^2 - |(\tilde{v}_j)_r|/\sqrt{2s}\right) = \varepsilon - \|(\tilde{v}_j)_S\|_2^2 + \|(\tilde{v}_j)_S\|_1/\sqrt{2s} \qquad \text{(B.3)}$$

$$\leq \varepsilon - (1 - \sqrt{t/2s})\|(\tilde{v}_j)_S\|_2^2 \ .$$

Here, the last inequality follows from the Cauchy-Schwarz inequality $\|(\tilde{v}_j)_S\|_1 \leq \sqrt{t} \cdot \|(\tilde{v}_j)_S\|_2$.

Now we use a simple proposition on the relationship between $\ell_1$ and $\ell_2$ norms: given $t \leq s/2$, $\ell_1$ norm $\|\tilde{v}_j\|_1 \leq \sqrt{s}/2$, and $\ell_2$ norm $\|\tilde{v}_j\|_2 = 1$, we must have $\|(\tilde{v}_j)_S\|_2 \geq \sqrt{t/s}$, i.e., much of the $\ell_2$ mass must lie on it $t$ largest coordinates (see Proposition B.3 below for its proof).

Combining this with (B.3) and $t/s > C\varepsilon$ gives:

$$\langle u_j, u_k \rangle \leq \varepsilon - \left(1 - \frac{1}{\sqrt{2}}\right) t/s < t/s \cdot (1/C - 2/C) = -\frac{t}{sC}$$

Now we can write

$$0 \leq \Big\|\sum_{j=1}^{N} u_j\Big\|_2^2 = \sum_{j=1}^{N} \|u_j\|_2^2 + \sum_{j \neq k} \langle u_j, u_k \rangle \leq N - \frac{tN(N-1)}{sC} \ ,$$

which gives $s \geq \frac{t(N-1)}{C}$ and completes the proof. $\square$

**Proposition B.3.** *Let $x \in \mathbb{R}^m$ with $\|x\|_1 \leq \sqrt{s}/2$ and $\|x\|_2 = 1$. Also, assume that $|x_1| \geq \ldots \geq |x_m|$. Then, for any $t \in [s/2]$, $\sum_{i=1}^{t} x_i^2 \geq \frac{t}{s}$ holds.*

*Proof.* Assume contrary: $\exists t \leq s/2 : \sum_{i=1}^{t} x_i^2 < \frac{t}{s}$. Then, since the absolute values of components are sorted, $x_t^2 < \frac{1}{s} \Rightarrow \forall j \geq t : x_j^2 < \frac{1}{s}$ and we have

$$\sqrt{\frac{1}{s}} \|x\|_1 > x_{t+1}^2 + \ldots + x_m^2 = 1 - \sum_{i=1}^{t} x_i^2 > 1 - \frac{t}{s} \geq \frac{1}{2}$$

However, this implies $\|x\|_1 > \frac{\sqrt{s}}{2}$, leading to a contradiction. $\square$

---

[22] The amount of $\ell_2$ mass contained in the top $t$ coordinates is at most $1 + t/(2s)$, so the sum of integer multiples of $1/(2s)$ that correspond to the rounded $t$ coordinates can be at most $2s + t$. Consider the representations of $2s + t$ as the sum of $t + 1$ non-negative integers. Then, each possible rounding has an unique representation, where the first $t$ summands correspond to the integer multiples of $1/(2s)$ and the last summand is the residual.

At last, we put in the right parameter of $t$ and conclude the proof. The following theorem resembles [NN13, Theorem 10].

**Theorem 3.** *There is some fixed $0 < \varepsilon_0 < 1/2$ so that the following holds. For any $1/\sqrt{n} < \varepsilon < \varepsilon_0$ and $m < O(n/\log(1/\varepsilon))$, let $A \in \mathbb{R}^{m \times n}$ be an $\varepsilon$-incoherent matrix. Then, at least half of the columns $A$ must have $\ell_1$ norm being $\Omega(\sqrt{\varepsilon^{-1} \log n / \log(m/\log n)})$.*

*Proof.* Let $s$ be a value such that half of $A$'s columns have $\ell_1$ norm at most $\sqrt{s}/2$, then we want to show that $s \geq \Omega(\varepsilon^{-1} \log n / \log(m/\log n))$.

By Lemma B.1, we have a weak lower bound $4\varepsilon s \geq 1$, allowing us to chose $t = 7\varepsilon s \geq 1$. We are now ready to prove that:
$$s \geq \frac{\log(7\varepsilon n/(4C))}{7\varepsilon \log\left(\frac{8e^2 m}{49\varepsilon^2 s}\right)} \;, \tag{B.4}$$
where $C$ is as in Lemma B.2. Assume contrary, then we get:
$$\left(\frac{8e^2 m}{49\varepsilon^2 s}\right)^{7\varepsilon s} < \frac{7\varepsilon n}{4C} \;.$$

Furthermore, for small enough $\varepsilon$,
$$2^{t+1}\binom{m}{t}\binom{2(s+t)}{t} \leq 2^{t+1} \frac{(em)^t}{t^t} \frac{(2e)^t (s+t)^t}{t^t} \leq 2 \cdot \left(\frac{8e^2 m}{49\varepsilon^2 s}\right)^{7\varepsilon s} < \frac{7\varepsilon n}{2C} \leq \frac{n}{2} \;,$$
so we can now apply Lemma B.2 and get:
$$\frac{sC}{t} \geq N - 1 \geq \frac{n}{2 \cdot 2^{t+1}\binom{m}{t}\binom{2(s+t)}{t}}.$$

By rearranging terms, it directly follows that
$$7\varepsilon n = \frac{tn}{s} \leq 2C \cdot 2^{t+1}\binom{m}{t}\binom{2(s+t)}{t} \leq 4C \cdot \left(\frac{8e^2 m}{49\varepsilon^2 s}\right)^{7\varepsilon s} < 7\varepsilon n \;,$$

giving a contradiction. This completes the proof of (B.4).

Let us now define $r = \log(7\varepsilon n/(4C))/(7\varepsilon)$ and $q = 8e^2 m/(49\varepsilon^2)$. Then we have $s \log(q/s) \geq r$ and for $\varepsilon < 1/2$, $q/e \geq m \geq s$. By [Alo09], $m = \Omega(\log n)$ and hence for small enough $\varepsilon$, $q/r > 2$ also holds. Using Proposition B.4 below, we get $s \geq \Omega(r/\log(q/r)) = \Omega(\varepsilon^{-1} \log n / \log(\varepsilon^{-1} m/\log n))$, since $\log(\varepsilon n) = \Theta(\log n)$ as $\varepsilon > 1/\sqrt{n}$. This is be equivalent to our theorem statement, since $m = \Omega(\frac{1}{\varepsilon})$ (using for instance the general lower bound on $m$ from [Alo09], or our weak sparsity lower bound Lemma B.1 as $m \geq \Omega(s)$). □

**Proposition B.4.** *Let $s, q, r$ be positive reals with $q \geq \max(2r, es)$. Then, if $s \log(q/s) \geq r$ it must be the case that $s = \Omega(r/\log(q/r))$.*

*Proof.* The function $f(s) = s \log(q/s)$ is non-decreasing for $s \leq q/e$ since $f'(s) = \log(q/(es)) \geq 0$. Since we are proving a lower bound on $s$, we can without the loss of generality consider $s \log(q/s) = r$. From here with $q/s \geq e$ immediately follows that $s \leq r$. , $r/s = \log(q/s) = \log(q/r) + \log(r/s)$.

Finally, we can write:
$$\frac{s}{r/\log(q/r)} = \frac{s((r/s) - \log(r/s))}{r} = 1 - \frac{s}{r}\log\left(\frac{r}{s}\right) \geq 1 - \frac{1}{e} \qquad □$$

# C Proof of Theorem 4: Lower Bounds for Non-Negative RIP Matrices

In this section we prove Theorem 4. The proof framework is inspired from that of Theorem 2: we will first prove a sparsity lower bound on RIP matrices, and then use the sparsity lower bound to deduce a dimension lower bound on $m$.

However, the $\ell_0$-sparsity lower bound of $\tilde{\Omega}(k \log(d/k))$ for general RIP matrices from [NN13] is not tight, so we cannot use it directly. We show, instead, for non-negative RIP matrices a tight (and even $\ell_1$-sparsity) lower bound can be deduced using very similar idea from the proof of our strengthened Theorem 3. We shall explain this in Appendix C.1.

The second part of the proof is simpler. It uses the same averaging idea from our Lemma 5.2, but makes use of the RIP property rather than the JL property. We will explain this in Appendix C.2.

## C.1 Sparsity Lower Bound for Non-Negative RIP Matrices

We begin with a simple fact on RIP matrices. This is analogous to for instance [NN13, Lemma 3], but is stronger since we are dealing with RIP matrices (which are also $\varepsilon$-incoherent matrices).

**Fact C.1.** *Let $A \in \mathbb{R}^{m \times d}$ be a non-negative $(k, \varepsilon)$-RIP matrix. For any $x \geq \frac{3\varepsilon}{k-1}$, we have that any row of $A$ must have no more than $\frac{3\varepsilon}{x} + 1$ entries of magnitude greater than or equal to $\sqrt{x}$.*

*Proof.* Suppose that $A$ has such a row, say the $r$-th row, with $N = \lceil \frac{3\varepsilon}{x} + 1 \rceil$ entries whose magnitudes are greater than $\sqrt{x}$. Note that we must have $N \leq k$ because $k \geq \frac{3\varepsilon}{x} + 1$ by our assumption on $x$. Let $\tilde{v}_1, \ldots, \tilde{v}_N \in \mathbb{R}^m$ denote the vectors for these $N$ columns, and because we have assumed that $A$ is column normalized, $\|\tilde{v}_1\|_2 = \cdots = \|\tilde{v}_N\|_2 = 1$. Moreover, for any $i, j \in [N]$, we have $\langle \tilde{v}_i, \tilde{v}_j \rangle \geq x$ because both of them have their $r$-th greater than $\sqrt{x}$, and the rest of the coordinates non-negative. As a consequence,

$$\|\tilde{v}_1 + \cdots + \tilde{v}_N\|_2^2 = \sum_i \|\tilde{v}_i\|_2^2 + \sum_{i \neq j} \langle \tilde{v}_i, \tilde{v}_j \rangle \geq N + N(N-1)x .$$

On the other hand, because $N \leq k$ and $A$ is an $(k, \varepsilon)$-RIP matrix, it satisfies that $\|\tilde{v}_1 + \cdots + \tilde{v}_N\|_2^2 \leq (1+\varepsilon)^2 N$, so we conclude with

$$(1+\varepsilon)^2 N \geq N + N(N-1)x \implies N \leq \frac{2\varepsilon + \varepsilon^2}{x} + 1 \leq \frac{3\varepsilon}{x} + 1 .$$

$\square$

We then prove a lemma analogous to Lemma B.1. This is a weak lower bound $s \geq \Omega(\frac{k}{\varepsilon})$ for the $\ell_1$-sparsity of $(k, \varepsilon)$-RIP matrices, but will play an important role in the final proof of the stronger lower bound: $s = \tilde{\Omega}(\frac{k \log(d/k)}{\varepsilon})$.

**Lemma C.2.** *Suppose $m < \frac{d}{4(3\varepsilon \log(k/3\varepsilon)+1)} = O(\frac{d}{\varepsilon \log(k/\varepsilon)})$, and $A \in \mathbb{R}^{m \times d}$ is non-negative $(k, \varepsilon)$-RIP matrix. If $A$ has $d/2$ columns with $\ell_1$ norm at most $\sqrt{s/2}$ each, then $s \geq \frac{k-1}{6\varepsilon} \geq \frac{k}{12\varepsilon}$.*

*Proof.* Let $x_0 = \frac{3\varepsilon}{k-1}$ and for the sake of contradiction, assume $s < \frac{k-1}{6\varepsilon} = \frac{1}{2x_0}$.

Let $W \subset [d]$ be a set of the $d/2$ columns each with $\ell_1$ norm at most $\sqrt{s/2}$. Define $L(W) = \{(i,j) : A_{i,j}^2 \geq x_0, v_j \in W\}$ to be the set of large entries, and $S(W) = \{(i,j) : A_{i,j}^2 < x_0, v_j \in W\}$ to be that of small entries. Clearly, $S(W)$ and $L(W)$ are disjoint and span all the entries in $W$. Let us bound the sum of squares of matrix entries in $S(W)$ and $L(W)$ respectively.

- Small entries in $W$ are less than $\sqrt{x_0}$ in magnitude, so their squares sum up to at most $\sqrt{x_0}$ times the $\ell_1$ norm of such entries:

$$\sum_{i,j\in S(W)} A_{i,j}^2 \leq \sqrt{x_0} \cdot \Big(\sum_{i,j\in S(W)} |A_{i,j}|\Big) = \sqrt{x_0} \cdot \Big(\sum_{i,j\in W} |A_{i,j}| - \sum_{i,j\in L(W)} |A_{i,j}|\Big)$$

$$\leq \sqrt{x_0} \cdot \Big(\sum_{i\in W} \|v_i\|_1 - \sqrt{x_0}|L(W)|\Big) \leq \frac{d}{2}\sqrt{x_0 s/2} - x_0|L(W)| < \frac{d}{4} - x_0|L(W)| \quad \text{(C.1)}$$

- For large entries, let $X$ denote the square of a random entry from $L(W)$. Then,

$$\sum_{i,j\in L(W)} A_{i,j}^2 = |L(W)|\cdot\mathbb{E}[X] = |L(W)|\cdot\int_{x=0}^{1} \Pr[X > x]dx \leq x_0|L(W)| + m\cdot\int_{x=x_0}^{1} \Big(\frac{3\varepsilon}{x}+1\Big)dx$$

$$= x_0|L(W)| + m(3\varepsilon\log(1/x_0) + 1) \leq x_0|L(W)| + \frac{d}{4} \ . \quad \text{(C.2)}$$

Here the first inequality is due Fact C.1: there cannot be more than $\frac{3\varepsilon}{x} + 1$ entries in each row of absolute value more than $x$, whenever $x \geq x_0$. The second inequality is owing to the choice of $m < \frac{d}{4(3\varepsilon\log(k/3\varepsilon)+1)}$.

Now we can combine equations (C.1) and (C.2) to get $\sum_{i,j\in S(W)\cup L(W)} A_{i,j}^2 < \frac{d}{2}$. On the other hand, $\sum_{i,j\in S(W)\cup L(W)} A_{i,j}^2 = \sum_{v_i\in W} \|v_i\|_2^2 = \frac{d}{2}$ and we get a contradiction. $\square$

The above lower bound is weak since it is obtained merely from a counting argument. Let us now strengthen it into a stronger form, by using the pigeon-hole principle to find $N$ columns that pairwisely and positively correlate to each other. This cannot happen if the original matrix is $(k,\varepsilon)$-RIP. Let us explain:

**Lemma C.3.** *Let $0 < \varepsilon < 1/2$, $A \in \mathbb{R}^{m\times d}$ be a non-negative $(k,\varepsilon)$-RIP matrix, and $s$ be any value such that half of $A$'s columns have $\ell_1$ norm at most $\sqrt{s}/2$. Define $C = 1/(1 - 1/\sqrt{2}) = 2 + \sqrt{2}$. Then, for any $t \in [s/2]$, we must have $s \geq \frac{t(\min\{k,N\}-1)}{3\varepsilon C}$ with $N = \Big\lceil \frac{d}{2\binom{m}{t}\binom{2(s+t)}{t}}\Big\rceil$.*

*Proof.* The proof structure almost the same as Lemma B.2: for any vector $v_i$, consider its $t$ largest coordinates in absolute magnitude, and define its $t$-type to be a pair containing:

- the locations of the top $t$ coordinates (there are at most $\binom{m}{t}$ choices);
- the rounding of the top $t$ values so that their squares round to the nearest integer multiple of $1/(2s)$. Values halfway between two multiples can be rounded arbitrarily. (There are at most $\binom{2s+2t}{t}$ number of different roundings, for the same reason as Lemma B.2.)

All in all, there are $\binom{m}{t}\binom{2s+2t}{t}$ possible $t$-types. By the pigeon-hole principle, out of $d/2$ column vectors that have $\ell_1$ norm at most $\sqrt{s}/2$, we can select $N$ vectors $\tilde{v}_1,\ldots,\tilde{v}_N$, such that they all have the same $t$-type.

Let $S \subset [d]$ be the set of the largest coordinates for these vectors, and we have $|S| = t$. Now, for $j \neq k \in [N]$, since $\tilde{v}_j$ and $\tilde{v}_k$ have the same type, we have

$$\langle \tilde{v}_j, \tilde{v}_k\rangle \geq \sum_{r\in S}(\tilde{v}_j)_r(\tilde{v}_k)_r \geq \sum_{r\in S}(\tilde{v}_j)_r\big((\tilde{v}_j)_r - 1/\sqrt{2s}\big)$$

$$= \sum_{r\in S}\Big((\tilde{v}_j)_r^2 - (\tilde{v}_j)_r/\sqrt{2s}\Big) = \|(\tilde{v}_j)_S\|_2^2 + \|(\tilde{v}_j)_S\|_1/\sqrt{2s} \quad \text{(C.3)}$$

$$\geq (1 - \sqrt{t/2s})\|(\tilde{v}_j)_S\|_2^2 \ .$$

Here, the last inequality follows from the Cauchy-Schwarz inequality $\|(\tilde{v}_j)_S\|_1 \leq \sqrt{t} \cdot \|(\tilde{v}_j)_S\|_2$.

Now we use again the relationship between $\ell_1$ and $\ell_2$ norms: given $t \leq s/2$, $\ell_1$ norm $\|\tilde{v}_j\|_1 \leq \sqrt{s}/2$, and $\ell_2$ norm $\|\tilde{v}_j\|_2 = 1$, we must have $\|(\tilde{v}_j)_S\|_2 \geq \sqrt{t/s}$, i.e., much of the $\ell_2$ mass must lie on it $t$ largest coordinates (see Proposition B.3).

Combining this with (C.3) gives:
$$\langle \tilde{v}_j, \tilde{v}_k \rangle \geq \left(1 - \frac{1}{\sqrt{2}}\right) t/s = \frac{t}{sC}$$

Now let $N' = \min\{k, N\}$ and we focus on the first $N'$ columns in $\{\tilde{v}_1, \ldots, \tilde{v}_N\}$. By the definition of $(k, \varepsilon)$-RIP, we have

$$(1 + 3\varepsilon)N' \geq (1 + \varepsilon)^2 N' \geq \Big\|\sum_{j=1}^{N'} \tilde{v}_i\Big\|_2^2 = \sum_{j=1}^{N'} \|\tilde{v}_j\|_2^2 + \sum_{j \neq k \in [N']} \langle \tilde{v}_j, \tilde{v}_k \rangle \geq N' + \frac{tN'(N'-1)}{sC} \ ,$$

which gives $s \geq \frac{t(N'-1)}{3\varepsilon C}$ and completes the proof. $\square$

We are now ready to choose the right parameter of $t$ and conclude the proof. The following theorem resembles Theorem 2.

**Theorem 4a.** *There is some fixed $0 < \varepsilon_0 < 1/2$ so that the following holds. For any $0 < \varepsilon < \varepsilon_0$ and $m < O(\frac{d}{\varepsilon \log(k/\varepsilon)})$, let $A \in \mathbb{R}^{m \times d}$ be a non-negative $(k, \varepsilon)$-RIP matrix. Then, at least half of the columns $A$ must have $\ell_1$ norm being $\Omega\Big(\sqrt{\frac{k \log(d/k)}{\varepsilon}} / \log\big(m/\log(d/k)\big)\Big)$.*

*Proof.* Let $s$ be a value such that half of $A$'s columns have $\ell_1$ norm at most $\sqrt{s}/2$, then we want to show that $s \geq \Omega(\varepsilon^{-1} \log d / \log(m/\log d))$.

By Lemma C.2, we have a weak lower bound $s \geq \frac{k}{12\varepsilon}$, allowing us to chose $t = \frac{6C\varepsilon s}{k} \geq 1$, where $C = 2 + \sqrt{2}$ is the constant from Lemma C.3. We are now ready to prove that:
$$s \geq \frac{k \log(d/4k)}{6C\varepsilon \log\left(\frac{e^2 mk^2}{36\varepsilon^2 s}\right)} \ , \tag{C.4}$$

where $C$ is as in Lemma B.2. Assume contrary, then we get:
$$\left(\frac{4e^2 mk^2}{144\varepsilon^2 s}\right)^{6C\varepsilon s/k} < \frac{d}{4k} \ .$$

Furthermore, we have
$$2\binom{m}{t}\binom{2(s+t)}{t} \leq 2 \frac{(em)^t}{t^t} \frac{(2e)^t (s+t)^t}{t^t} \leq 2 \cdot \left(\frac{4e^2 mk^2}{144\varepsilon^2 s}\right)^{6C\varepsilon s/k} < \frac{d}{4k} \ ,$$

and this allows us to apply Lemma C.3 with $N \geq k$ (while noticing that for sufficiently small $\varepsilon$ we have $t \in [s/2]$):
$$s \geq \frac{t(k-1)}{3\varepsilon C} = \frac{\frac{6C\varepsilon s}{k}(k-1)}{3\varepsilon C} > s \ ,$$
giving a contradiction. This completes the proof of (C.4).

Let us now define $r = \frac{k}{6C\varepsilon} \log \frac{d}{4k}$ and $q = \frac{e^2 mk^2}{36\varepsilon^2}$. Then we have $s \log(q/s) \geq r$ and for sufficiently small $\varepsilon < 1/2$, $q/e \geq m \geq s$. By [Alo09], $m = \Omega(\log d)$ and hence for small enough $\varepsilon$, $q/r > 2$ also holds. Using Proposition B.4, we get $s \geq \Omega(r/\log(q/r)) = \Omega\Big(\frac{k \log(d/k)}{\varepsilon} \cdot \frac{1}{\log\big(\varepsilon^{-1} mk \log^{-1}(d/k)\big)}\Big)$.
At last, using the fact that $m \geq \Omega(k/\varepsilon)$ (for instance by our Lemma C.2 and $m \geq \Omega(s)$), this is equivalent to our theorem statement. $\square$

## C.2 Dimension Lower Bound for Non-Negative RIP Matrices

Now we want to turn the $\ell_1$-sparsity lower bound into dimension lower bound for RIP matrices. The proof below reuses the idea we sketched in Section 5.2, but is much simpler than the JL case.

Essentially, we first use the $\ell_1$-sparsity lower bound on $A$ to deduce that $A$ must have a large average column correlation. Namely, $\sum_{i,j} \langle v_i, v_j \rangle$ is large where $v_i$ represents the $i$-th column of $A$. By an averaging argument, we can pick some subset $S \subset [d]$ of the columns where $|S| = k$, such that the correlations between columns in $S$ are also large: namely, $\sum_{i,j \in S} \langle v_i, v_j \rangle \geq \Omega(sk^2/m)$.

Next, the property of RIP, $A$ must satisfy that $\|\sum_i v_i\|_2^2 = (1 \pm \varepsilon)k$ because $A$ preserves the $\ell_2$-norm for the $k$-sparse vector $x = \sum_{i \in S} e_i$. However, since the columns within $S$ positively correlate pairwisely by a large factor, this will contradict the fact that $\|\sum_i v_i\|_2^2 = k + \sum_{i,j \in S} \langle v_i, v_j \rangle \leq (1 + O(\varepsilon))k$, unless $m$ is large.

**Theorem 4b.** *There is some fixed $0 < \varepsilon_0 < 1/2$ so that the following holds. For any $0 < \varepsilon < \varepsilon_0$ and $m < O(\frac{d}{\varepsilon \log(k/\varepsilon)})$, let $A \in \mathbb{R}^{m \times d}$ be a non-negative $(k, \varepsilon)$-RIP matrix. Then, we must have $m = \Omega\left(\frac{k^2 \log(d/k)}{\varepsilon^2} / \log\left(\varepsilon^{-2} k / \log(d/k)\right)\right)$.*

*Proof.* Let $s$ be a value such that half of $A$'s columns have $\ell_1$ norm at least $\sqrt{s}$, and according to Theorem 4a we have $s = \Omega\left(\frac{k \log(d/k)}{\varepsilon} / \log\left(m / \log(d/k)\right)\right)$. Without loss of generality let them be the first $d/2$ columns: $v_1, \ldots, v_{d/2} \in \mathbb{R}^m$.

Next, let us compute the pairwise inner products between such vectors. We have

$$\sum_{\substack{i,j \in [d/2] \\ i \neq j}} \langle v_i, v_j \rangle = \sum_{r \in [m]} \Big( \sum_{i \in [d/2]} A_{r,i} \Big)^2 - \sum_{r \in [m]} \sum_{i \in [d/2]} A_{r,i}^2 \geq \frac{1}{m} \Big( \sum_{r \in [m]} \sum_{i \in [d/2]} A_{r,i} \Big)^2 - \frac{d}{2} \geq \frac{d^2 s}{4m} - \frac{d}{2} \geq \frac{d^2 s}{5m}$$

where the last inequality uses the fact that $m \ll d$ and $s \geq 1$. We now try to conduct an averaging argument (like we did in Lemma 5.2). If we consider all subset $S \subset [d/2]$ of the columns with cardinality $k$, their average pairwise inner products must be be large, so there exists one of them with good pairwise inner products. Let us explain. We begin with

$$\sum_{\substack{S \subset [d/2] \\ |S| = k}} \sum_{\substack{i,j \in S \\ i \neq j}} \langle v_i, v_j \rangle = \binom{d/2 - 2}{k - 2} \cdot \sum_{\substack{i,j \in [d/2] \\ i \neq j}} \langle v_i, v_j \rangle \ ,$$

and note that there are a total number of $\binom{d/2}{k}$ subsets $S$ of cardinality $k$. By an averaging argument, there exist some subset $S^* \subset [d/2]$ satisfying

$$\sum_{\substack{i,j \in S^* \\ i \neq j}} \langle v_i, v_j \rangle \geq \frac{1}{\binom{d/2}{k}} \binom{d/2 - 2}{k - 2} \cdot \sum_{\substack{i,j \in [d] \\ i \neq j}} \langle v_i, v_j \rangle \geq \Omega\Big(\frac{k^2}{d^2}\Big) \cdot \frac{d^2 s}{5m} = \Omega\Big(\frac{sk^2}{m}\Big) \ .$$

Without loss of generality, we assume that $S^* = \{1, 2, \ldots, k\}$ are the first $k$ columns of $A$. By the $(k, \varepsilon)$-RIP property of $A$, the $\varepsilon$-distortion of the $k$-sparse vector $x = e_1 + \cdots + e_k$ yields

$$(1 + \varepsilon)^2 \|x\|_2^2 = (1 + \varepsilon)^2 k \geq \|Ax\|_2^2 = \Big\| \sum_{i=1}^k v_i \Big\|_2^2 = k + \Omega\Big(\frac{sk^2}{m}\Big) \ ,$$

and this immediately implies $m \geq \Omega(\frac{sk}{\varepsilon})$. Combining this with $s = \Omega\left(\frac{k \log(d/k)}{\varepsilon} / \log\left(m / \log(d/k)\right)\right)$, we arrive at our desired lower bound $m = \Omega\left(\frac{k^2 \log(d/k)}{\varepsilon^2} / \log\left(\varepsilon^{-2} k / \log(d/k)\right)\right)$. □

# D  Simple Lower Bound for Non-Negative JL Matrices

In this section we show a simple fact: at least in the interesting parameter regime of $\delta = 1/\mathsf{poly}(d)$, we must have $m \geq \Omega(d)$ in order to construct a non-negative JL matrix. Since we cannot find the proof of this simple fact anywhere else, we provide it below.

**Fact D.1.** *Let $\mathcal{A}$ be a distribution over $m \times d$ non-negative matrices such that, for any $x \in \mathbb{R}^d$, with probability at least $1 - \delta$, the $\ell_2$ embedding $\|Ax\|_2 = (1 \pm \varepsilon)\|x\|_2$ has $\varepsilon$-distortion. Then,*

$$m \geq (1 - 4\varepsilon) \min\left\{d, \frac{1}{\delta} - 2\right\} \; .$$

*Proof.* Given any such distribution $\mathcal{A}$, we choose $n = \min\{d, \frac{1}{\delta} - 2\}$. Using union bound, with probability at least $1 - \delta(n+1) > 0$, a matrix $A$ drawn from $\mathcal{A}$ will preserve $\ell_2$ norms with $\varepsilon$ distortion for all vector $x \in \{e_1, \ldots, e_n\} \cup \{e_1 + e_2 + \cdots + e_n\}$.

This implies that, the $\ell_2$-norm of each of the first $n$ columns of $A$ is at least $1-\varepsilon$: this is because for every $j \in [n]$, $\sqrt{\sum_{i \in [m]} A_{i,j}^2} = \|Ae_j\|_2 \geq (1-\varepsilon)\|e_j\|_2 = 1 - \varepsilon$.

Next, we check the norm preservation on $x = e_1 + e_2 + \cdots + e_n \in \mathbb{R}^d$. Its $\ell_2$ norm is $\|x\|_2 = \sqrt{n}$, so we must have $\|Ax\|_2^2 \leq n(1+\varepsilon)^2$. On the other hand,

$$\|Ax\|_2^2 = \sum_{i=1}^m \left(\sum_{j=1}^n A_{i,j}\right)^2 \geq \frac{1}{m}\left(\sum_{i=1}^m \sum_{j=1}^n A_{i,j}\right)^2 \geq \frac{1}{m}\left(\sum_{j=1}^n \|Ae_j\|_1\right)^2 \geq \frac{1}{m}\left(\sum_{j=1}^n \|Ae_j\|_2\right)^2$$
$$\geq \frac{1}{m}\left((1-\varepsilon)n\right)^2 \; .$$

Together, they imply $m \geq (1 - 4\varepsilon)n$. $\square$